\documentclass[fleqn,usenatbib]{mnras}

\usepackage{newtxtext,newtxmath}
\usepackage[T1]{fontenc}

\DeclareRobustCommand{\VAN}[3]{#2}
\let\VANthebibliography\thebibliography
\def\thebibliography{\DeclareRobustCommand{\VAN}[3]{##3}\VANthebibliography}

\usepackage{graphicx}	% Including figure files
\usepackage{amsmath}	% Advanced maths commands
\usepackage{epstopdf}
\usepackage{enumerate}
\usepackage{soul}
\usepackage{savesym}
\savesymbol{tablenum}
\usepackage{siunitx}
\usepackage{cleveref}
\restoresymbol{SIX}{tablenum}
\usepackage{hyperref}

\newcommand{\del}{\partial}

\title{A formation scenario of black hole-envelope systems --viscous hydrodynamics simulation in general relativity--}% Force line breaks with \\

\author[Alan Tsz Lok Lam et al.]{
Alan Tsz Lok Lam,$^{1,2,3}$\thanks{E-mail: tpl5641@psu.edu}
Masaru Shibata,$^{3,4}$
Kenta Hotokezaka$^{5,3}$
Carlo Musolino$^{3}$
\\
$^{1}$Institute for Gravitation and the Cosmos, The Pennsylvania State University, University Park, PA 16802, USA\\
$^{2}$Department of Physics, The Pennsylvania State University, University Park, PA 16802, USA\\
$^{3}$Max-Planck-Institut f\"ur Gravitationsphysik (Albert-Einstein-Institut), Am M\"uhlenberg 1, D-14476 Potsdam-Golm, Germany\\
$^{4}$Center for Gravitational Physics and Quantum Information, Yukawa Institute for Theoretical Physics, Kyoto University, Kyoto, 606-8502, Japan\\
$^{5}$Research Center for the Early Universe, Graduate School of Science, The University of Tokyo, Bunkyo, Tokyo 113-0033, Japan\\
}

\date{Accepted XXX. Received YYY; in original form ZZZ}

\begin{document}
\label{firstpage}
\pagerange{\pageref{firstpage}--\pageref{lastpage}}
\maketitle

\begin{abstract}
By performing a viscous hydrodynamics simulation in general relativity for super-Eddington accretion flows onto massive black holes of mass $M=10^5$--$10^7M_\odot$, we discuss a formation scenario for black hole-envelope systems. We consider the mass accretion rate of $a^3/G \approx 1.5 \times 10^{25} (a/10\,\mathrm{km\,s^{-1}})^3$\,g/s, comparable to the Eddington mass accretion rate of a $10^7M_\odot$ black hole, assuming that the gas temperature of the infalling matter is $\lesssim 10^4$\,K. Here, $a$ and $G$ denote the sound speed and gravitational constant. For the accretion flow, we set up a quasi-spherical Bondi-type flow in which radial inflow dominates over angular momentum in the distant region. It is found that (i) for low-mass black holes with $M \lesssim 10^6M_\odot$, a photon-trapped region forms in the inner region, and a significant viscous outflow driven near the polar region overcomes the ram pressure of the mass inflow, leading to an inflow-outflow structure; (ii) for massive black holes of $M \gtrsim 3 \times 10^6M_\odot$, the outflow is not launched, and a convective envelope around the black hole gradually develops; and (iii) irrespective of the black-hole mass, the mass accretion rate onto the black hole is of order 10\% of the Eddington accretion rate for reasonable values of the viscous coefficient. As the mass accretion rate onto the black holes is much lower than the mass growth rate of the envelope for low-mass black holes with $M\lesssim 10^6M_\odot$, the envelope mass is likely to increase until the total viscous heating rate exceeds the Eddington luminosity of the system, if the mass accretion rate is preserved to be high for $\gtrsim 10^8 (M/10^7M_\odot)$\,yrs. 
\end{abstract} 

\begin{keywords}
gravitation – hydrodynamics – relativistic processes – stars: massive – stars: rotation - black hole
\end{keywords}

%\maketitle
\section{Introduction}

JWST observations have discovered a population of compact, luminous, high-redshift objects with a unique V-shaped spectrum referred to as little red dots (LRDs)~\citep{2023ApJ...959...39H, 2024ApJ...963..129M, 2024ApJ...964...39G, 2025ApJ...986..126K}. They are bright in UV and optical bands but faint in X-ray and radio bands~\citep{2024ApJ...974L..26Y, 2025MNRAS.538.1921M, 2026A&A...706A.372M}. Their compactness and broad emission lines suggest the presence of supermassive black holes (SMBHs) at their centers, with a high accretion rate of gas, like active galactic nuclei, although their faint X-ray and radio emission and low variability distinguish them from typical active galactic nuclei. 
 
The estimated masses of the SMBHs in LRDs are $10^6$--$10^8M_\odot$, lower than those of the most massive high redshift SMBHs (see, e.g., \citealt{2023ApJ...959...39H, 2024ApJ...963..129M, 2025arXiv250821748J, 2025arXiv251107515K, 2026arXiv260118864S}). These mass estimates for LRDs, however, remain debated (see e.g., \citealt{2026Natur.649..574R, 2025MNRAS.544L.167B}). A widely discussed interpretation of such LRDs is that they are growing black holes embedded in dense gas, although the total gas mass has a wide variety depending on the scenarios, ranging from low-mass cocoon-like structures to quasistar-like massive hydrostatic envelopes (see, e.g.,~\citealt{Begelman:2006db, Begelman:2007je, 2025arXiv250316596N, 2025ApJ...980L..27I, 2025MNRAS.544.3407K, 2025arXiv251121820D}).

Assuming that the gas accretion rate onto SMBHs depends broadly on the virial gas temperature $T$ in the central region of the galaxies, the mass accretion rate $\dot M$ can be approximated as $a^3/G$ (e.g., ~\citealt{1977ApJ...214..488S, Begelman:2006db, 2016MNRAS.459.3738I}), where $G$ is the gravitational constant and $a$ is the sound speed $\propto T^{1/2}$. Here, $T$ denotes the gas temperature with $T\lesssim 10^4$\,K, i.e., $a \sim 10$\,km/s, leading to $\dot M \sim 10^{25}$ g/s, similar to the Eddington accretion rate for SMBHs with mass $M \sim 10^7M_\odot$. Thus, for $M \lesssim 10^7M_\odot$, the accretion is likely super-Eddington if $T\sim 10^4$ K. In such super-Eddington cases, the gas density in the vicinity of SMBHs is high enough that photons are trapped in the accretion flow. In the presence of viscosity (or effective viscous effects from magnetohydrodynamic turbulence), the dissipation of kinetic energy produces substantial thermal energy, which can drive convection, outflows, and jets (e.g.,~\citealt{2014MNRAS.439..503S, 2014MNRAS.441.3177M, 2016MNRAS.456.3929S, 2018PASJ...70..108K, 2021PASJ...73..450K, 2022PASJ...74.1378Y, 2024MNRAS.532.4826T, Shibata:2025ycs}). 

In this paper, we report the results of our viscous hydrodynamics simulations in general relativity. We focus on the (shear) viscous evolution of quasi-spherical accretion flows onto SMBHs with $M \leq 10^7M_\odot$ at a fixed mass accretion rate at the sonic point of $\approx 1.5 \times 10^{25}$\,g/s, comparable to the Eddington mass accretion rate for $10^7M_\odot$ SMBHs. We examine the effects of the central engine (i.e., viscous heating in the formed torus) on the entire evolution of the system by accurately resolving the central region around the black hole, while employing a simplified equation of state for the matter field. The simulations are performed over a long timescale with $t >3 \times 10^{6} GM/c^3$, where $c$ is the speed of light. 
In this paper, we will illustrate that for super-Eddington accretion with $M \ll 10^7M_\odot$, an outflow is driven by viscous dissipation within the torus formed near the black hole if the specific angular momentum is sufficiently large to allow torus formation. By contrast, for near-Eddington accretion, i.e., for $M \gtrsim 10^{6.5}M_\odot$, no outflow is driven, and a torus gradually grows as matter accretes; a quasi-steady black hole-envelope formation is found. We also show that the growth rate of the black hole is always lower than that of the envelope for the black hole masses considered in this paper; the envelope mass is likely to increase until the total viscous heating rate becomes comparable to the Eddington luminosity of the system if the mass accretion continues for a sufficiently long timescale.

The paper is organized as follows. In Sec.~\ref{sec2}, we describe the model setup. Section~\ref{sec3} presents numerical results from viscous hydrodynamics simulations across a range of SMBH masses and inflow specific angular momentum. After discussing the implications of the numerical results in Sec.~\ref{sec4}, we summarize the paper in Sec.~\ref{sec5}. 
Throughout this paper, we use the geometrical units with $G=1=c$ unless otherwise stated. For the units of length and time, we use $r_\mathrm{g}=GM/c^2$ and $t_\mathrm{g}=r_\mathrm{g}/c$, respectively.

\section{Model}\label{sec2}

In this paper, we consider quasi-spherical, axisymmetric accretion flows onto a non-spinning black hole of mass $M$. Specifically, the flow is assumed to be nearly spherical at large radii ($r \gg r_\mathrm{g}$), even though a small specific angular momentum $j$ is included.
In this setting, it is natural to choose the Bondi-type flow as the background flow in the limit $j \rightarrow 0$. Thus, we first review the structure of the Bondi flow under spherical symmetry and without viscosity. For the present work, the mass of the accreting matter is assumed to be much smaller than $M$ so that the self-gravity of the matter can be ignored.

In the Bondi solution \citep{1952MNRAS.112..195B, 1983bhwd.book.....S}, the flow is subsonic outside the sonic point of $r=r_\mathrm{s}$ and becomes supersonic for $r < r_\mathrm{s}$. We denote the rest-mass density $\rho$ and the radial velocity $u = -u^r (>0)$ at the sonic point as $\rho_\mathrm{s}$ and $u_\mathrm{s}$, respectively, in the following. Here, $u^\mu$ denotes the four velocity and $u_\mathrm{s}$ is given by $\sqrt{M/(2r_\mathrm{s})}$. 

The basic equations for the Bondi solution are the continuity equation and the Euler equation. Under steady-state assumptions, the former leads to 
\begin{equation}
\dot M:=4\pi \rho u r^2=4\pi \rho_\mathrm{s} u_\mathrm{s} r_\mathrm{s}^2=\mathrm{const}., \label{eq2}
\end{equation}
where $\dot M$ denotes the rest-mass injection rate, and later, under the assumption of the adiabatic flow, leads to  
\begin{equation}
h^2 \left(1 - \frac{2M}{r} + u^2\right)=h_\mathrm{s}^2
\left(1- \frac{3M}{2r_\mathrm{s}}\right), \label{eq3}
\end{equation}
where $h$ denotes the specific enthalpy and $h_\mathrm{s}$ is $h$ evaluated at $r=r_\mathrm{s}$. Here, we assumed the use of Schwarzschild or Eddington-Finkelstein coordinates~(e.g., \citealt{Hawking:1973uf}). Even in the presence of cooling, the Euler equation can be integrated for the isothermal flow, and in this case, equation~\eqref{eq3} is replaced by
\begin{equation}
\left(1 - \frac{2M}{r} + u^2\right)(u r^2)^{-2a^2}=
\left(1- \frac{3M}{2r_\mathrm{s}}\right)(u_\mathrm{s} r_\mathrm{s}^2)^{-2a^2}, \label{eq4}
\end{equation}
where $a$ denotes sound speed (see below). 

For a given equation of state, equations~\eqref{eq2} and \eqref{eq3} constitute a system of simultaneous equations for $\rho$ and $u$, and the solutions are straightforward to obtain. In this paper, we consider the case in which the flow is nearly isothermal in a distant region. Specifically, we assume that the gas temperature in the isothermal region is $T \lesssim 10^4$\,K, and that the gas is composed of the primordial gas with a hydrogen-to-helium mass ratio of 3:1. For simplicity, the gas is assumed to be fully ionized. Then, the equation of state for the gas is written as
\begin{equation}
P_\mathrm{g}=\frac{Y_T k T}{m} \rho,~~\varepsilon_\mathrm{g}=\frac{3 Y_TkT}{2 m},\label{eq5}
\end{equation}
where $P$ is the pressure, $k$ is the Boltzmann constant, $Y_T \approx 27/16$, and $m$ is the atomic mass unit. In this equation of state, the sound speed may be defined as $a^2 \approx Y_T kT/m$ for $a \ll c$. 

As shown below, the photosphere lies within the inner region for the mass accretion rates considered in this paper. In the very innermost region, photons are trapped, and the photon radiation pressure, $a_\mathrm{r}T^4/3$, where $a_\mathrm{r}$ is the radiation constant, greatly exceeds the gas pressure. Under these conditions, the equation of state effectively differs from equation~\eqref{eq5}. This will be explained in Sec.~\ref{sec3}. 

Assuming that the temperature near the sonic point is $\approx 7000$\,K, the sound speed is $\approx 10$\,km/s, which is significantly slower than the speed of light. Therefore, the location of the sonic point can be estimated as 
\begin{equation}
r_\mathrm{s}\approx \frac{M}{2a^2} \approx 4.49 \times 10^8 a_{6}^{-2} r_\mathrm{g},
\end{equation}
where $a_6=a/(10^6\,\mathrm{cm/s})$. 

For $r \ll r_\mathrm{s}$, regardless of the equation of state, the Bondi solution approximately satisfies \citep{1983bhwd.book.....S}
\begin{equation}\label{eq:profile}
 u \approx \sqrt{\frac{2M}{r}}, ~~~\rho \approx \frac{1}{2}\rho_\mathrm{s} 
 \left(\frac{r}{r_\mathrm{s}}\right)^{-3/2}. 
\end{equation}
We then approximately define the rest mass within the sonic point by
\begin{equation}
M_\mathrm{s}:= \int_0^{r_\mathrm{s}} 4 \pi \rho r^2 dr \approx \frac{4 \pi} {3}\rho_\mathrm{s} r_\mathrm{s}^3.\label{eq8}
\end{equation}
Because we assume that the self-gravity of the matter is negligible compared with the black hole's gravity, we impose the condition $M_\mathrm{s} < M$.

From equation~\eqref{eq2}, $\rho_\mathrm{s}$ is written as
\begin{equation}
 \rho_\mathrm{s}=
 \frac{\dot M}{4\pi}\left(\frac{2}{M r_\mathrm{s}^3}\right)^{1/2}.   
\end{equation}
Assuming that $\dot M=f a^3 /G$ \citep{1977ApJ...214..488S} (we recover $G$ to clarify the unit) where $f$ is a constant, this equation is written to
\begin{equation}
\rho_\mathrm{s}=f \frac{M}{8\pi r_\mathrm{s}^3}.
\end{equation}
Thus if $f < 6$, the relation of $M_\mathrm{s} < M$ is satisfied. In the following numerical work, we assume $f=1$ for simplicity. In this case,
\begin{equation}
\rho_\mathrm{s}\approx 2.71 \times 10^{-22} a_6^6 M_6^{-2}\,\mathrm{g/cm^3},
\end{equation}
where $M_6=M/(10^6M_\odot)$, and hence, the gas pressure is about 11 orders of magnitude smaller than $a_\mathrm{r}T^4/3$ at $r=r_\mathrm{s}$ for $a_6=1=M_6$. If the flow remains entirely isothermal, the gas pressure can be comparable to the radiation pressure in the vicinity of the black hole. However, when shocks or viscous heating raise the temperature to $\gtrsim 10^5$ K, radiation pressure becomes always dominant.  

Assuming that the opacity is determined by the electron scattering of fully ionized gas, i.e., $\kappa \approx 0.35\,\mathrm{cm^2/g}$, the optical depth can be estimated by
\begin{equation} \label{eq:tau_est}
\tau=-\int^r_\infty \rho \kappa dr \approx \rho_\mathrm{s} \kappa r_\mathrm{s}^{3/2} r^{-1/2}.
\end{equation}
Then a photosphere, defined by $\tau=1$, is determined as
\begin{align}\label{eq:r_ps}
r_\mathrm{ps}&= (\rho_\mathrm{s} \kappa r_\mathrm{s})^2 r_\mathrm{s} 
\approx 1.78 \times 10^4 r_\mathrm{g} a_6^6 \kappa_{0.35}^2 M_6^{-2}f^2,
\end{align}
where $\kappa_{0.35}=\kappa/(0.35\,\mathrm{cm^2/g})$. Thus, the photosphere is formed inside the sonic point only for $M \gtrsim 10^4M_\odot$ with $f=1$. 

We can also define the photon-trapped region by $t_\mathrm{ff} < t_\mathrm{diff}$ where $t_\mathrm{ff}$ and $t_\mathrm{diff}$ denote the free-fall timescale of the flow and the diffusion timescale of photons. We simply define $t_\mathrm{ff}$ by
\begin{equation}
t_\mathrm{ff}:=\frac{r}{u} \approx \sqrt{{r^3 \over 2M}}.
\end{equation}
The diffusion timescale is defined in terms of the mean free path of photons, $\lambda \approx (\rho \kappa)^{-1}$, by
\begin{equation}
t_\mathrm{diff}:={3r^2 \over 4\lambda c}\approx {3 \over 4}\rho \kappa r^2 c^{-1}
\approx {3 \over 8}\rho_\mathrm{s} \kappa r_\mathrm{s}^{3/2} r^{1/2} c^{-1}.
\end{equation}
From $t_\mathrm{ff}=t_\mathrm{diff}$, we define the trapped radius by
\begin{align}
 r_\mathrm{trap} &={3 \over 4\sqrt{2}}r_\mathrm{g} \times (\rho_\mathrm{s} \kappa r_\mathrm{s})\sqrt{{r_\mathrm{s} \over r_\mathrm{g}}}={3 \over 4\sqrt{2}} r_\mathrm{g}\sqrt{{r_\mathrm{ps} \over r_\mathrm{g}}}
 \nonumber \\
 &\approx 71 r_\mathrm{g} a_6^3 \kappa_{0.35}M_6^{-1} f.\label{eq15}
\end{align}
Thus, the trapped region exists only for $M \lesssim 10^7M_\odot$ and is wider for lower-mass black holes. For $r \lesssim r_\mathrm{trap}$, the flow is adiabatic, with pressure governed by photon radiation, while for $r_\mathrm{trap} \lesssim r \lesssim r_\mathrm{ps}$, cooling by photon diffusion becomes important.
When cooling is highly effective, the assumption of fully ionized gas no longer holds and $\kappa$ should be smaller. In this scenario, both $r_\mathrm{ps}$ and $r_\mathrm{trap}$ decrease, resulting in a narrower adiabatic region. 

For $\dot M=a^3/G$, the accretion rate is written as
\begin{equation}
\dot M={a^3 \over G} \approx 1.50 \times 10^{25} a_6^{3} \,\,\mathrm{g/s}.
\label{eq16}
\end{equation}
On the other hand, the Eddington luminosity is written as
\begin{equation}
L_\mathrm{Edd}:={4 \pi cG M \over \kappa}=1.43 \times 10^{44} M_6 \kappa_{0.35}^{-1}\,\,\mathrm{erg/s}. \label{eq17}
\end{equation}
This gives an approximate Eddington mass accretion rate, defined by $L_\mathrm{Edd}/(0.1 c^2)$ (assuming that the viscous dissipation most efficiently occurs near the black hole), as
\begin{equation}
\dot M_\mathrm{Edd} \approx 1.59 \times 10^{24}M_6 \kappa_{0.35}^{-1}\,\,\mathrm{g/s}.\label{eq18}
\end{equation}
Therefore, super-Eddington accretion occurs only for $M \lesssim 10^7 M_\odot$. This is consistent with the fact that the trapped region appears only for $M \lesssim 10^7 M_\odot$. Moreover, equation~\eqref{eq18} indicates that the black hole growth timescale is 
\begin{equation}
t_\mathrm{Edd}:={M \over \dot M_\mathrm{Edd}}\approx 3.96 \times 10^7\,\kappa_{0.35}\,\,\mathrm{yrs}, \label{eq19}
\end{equation}
if the mass infall rate onto the black holes is equal to $\dot M_\mathrm{Edd}$.

In this paper, we focus on the accretion flow that has both the photosphere and a trapped region inside the sonic point. This condition is satisfied for accretion flows with $10^4M_\odot \lesssim M \lesssim 10^7M_\odot$. We also need to be careful about low-mass black holes because the radiation pressure near the photosphere may exceed the Eddington luminosity. Assuming a black-body photosphere at $T=7000$\,K, the black-body luminosity is written as
\begin{align}
L_\mathrm{bb@ps}&:=4 \pi r_\mathrm{ps}^2 \sigma T^4  \nonumber \\
&\approx 1.2 \times 10^{43} M_6^{-2} a_6^{12}\kappa_{0.35}^4 f^4 T_{7000}^4\,\mathrm{erg/s},
\end{align}
where $\sigma$ denotes the Stefan-Boltzmann constant and $T_{7000}=T/(7000\,\mathrm{K})$. It is found that $L_\mathrm{bb@ps}$ is larger than $L_\mathrm{Edd}$ for $M_6 \lesssim 0.4$ with $f=1$.
%As we already mentioned above, the free fall timescale is longer than the diffusion timescale at the photosphere. Thus, the radiation pressure is not directly reflected by $L_\mathrm{bb@ps}$ and may be lower than the Eddington value even for $M_6 \lesssim 0.4$ (note that the black body luminosity at $r=r_\mathrm{trap}$ is much lower than the Eddington luminosity). However, 
The large value of $L_\mathrm{bb@ps}$ suggests that the simple setting in this paper might not be justified for low-mass black holes, e.g., with $M \lesssim 10^5M_\odot$ (e.g., \citealt{2016MNRAS.459.3738I}). Therefore, in the following, we focus only on the case of $10^5M_\odot \leq M \leq 10^7M_\odot$. 

\section{Numerical simulation}\label{sec3}

We perform a viscous hydrodynamics simulation in general relativity for axisymmetric accretion flows around non-spinning black holes, adopting a fixed black-hole spacetime described by Eddington-Finkelstein coordinates $(t, r, \theta, \varphi)$ following \citet{1998ApJ...507L..67F}. With this choice, $\Omega^{-1}$, which is necessary to give the viscous coefficient (cf.~equation~\eqref{eq27}), is simply written as $(R^3/M)^{1/2}$, where $R=r\sin\theta$ denotes the cylindrical radius. 

Treating the Bondi flow described in Sec.~\ref{sec2} as a zeroth-order flow, we analyze a flow with specific angular momentum $j$, defined by
\begin{equation}
j=j_0 \sin^{2n}\theta,
\end{equation}
where $j_0$ is a constant and $n$ is a positive integer. This paper focuses on the case of $n=1$, since for $n > 1$, a larger fraction of the infalling matter simply directly accretes into the black hole. The initial conditions for $\rho$ and $u^r$ are based on the Bondi solution, with the additional condition $h u_\varphi=j$ while $u^\theta=0$ is unchanged. In the following, we describe the equation of state in Sec.~\ref{sec3.1}, the numerical setup in Sec.~\ref{sec3.2}, and present the numerical results in Sec.~\ref{sec3.3}. 

\subsection{Modeling equation of state}\label{sec3.1}

Accurately addressing this problem requires a radiation-transfer simulation together with evolving the ionization rate, as radiation cooling and matter opacity are crucial to the system's evolution. However, such simulations are computationally intensive, especially when a long-term evolution far beyond $10^6t_\mathrm{g}$ is needed, along with detailed resolution of viscous hydrodynamics near the black hole. To manage resources effectively, we instead utilize an equation of state that phenomenologically incorporates the cooling effect. 

Specifically, we write a hybrid equation of state in the form:
\begin{equation}
P=P_0(\rho) + (\Gamma_\mathrm{th}-1)\rho (\varepsilon-\varepsilon_0),
\end{equation}
where $P_0$ and $\varepsilon_0$ are written as
\begin{align}
P_0&=K\rho^\Gamma,\\
\varepsilon_0&=K (\Gamma-1)^{-1} \rho^{\Gamma-1},
\end{align}
and $K$ and $\Gamma$ are constants. 
Note that $P_0$ and $\varepsilon_0$ satisfy the thermodynamic relation $d\varepsilon_0=-P_0d\rho^{-1}$.
In this setting, in the absence of shock heating (and cooling), the polytropic form is preserved. As the flow is expected to be isothermal far from the black hole, we adopt a nearly isothermal equation of state in this region, setting $\Gamma=1.01$ and $a^2 \approx K$ with $a=10^6$ cm/s. 

$\Gamma_\mathrm{th}$ is chosen based on the flow conditions. In the trapped region, radiation pressure greatly exceeds gas pressure, so the equation of state is well approximated by $\Gamma_\mathrm{th}=4/3$. Conversely, in the optically thin region with $\tau < 1$, the isothermal condition should hold approximately (assuming an efficient cooling), leading us to set $\Gamma_\mathrm{th}=\Gamma=1.01$ for simplicity. For the intermediate region, we approximate $\Gamma_\mathrm{th}$ phenomenologically as
\begin{equation}
    \Gamma_\mathrm{th}={4 \over 3}-\left({\tau_\mathrm{tp}-\tau \over \tau_\mathrm{tp}-1}\right)^\xi\left({4 \over 3}-1.01\right). \label{eq24}
\end{equation}
Here, $\tau_\mathrm{tp}$ represents an approximate optical depth value at the outer boundary of the trapped region, with $\xi$ being a constant. In our analysis, we set $\tau_\mathrm{tp}=15 M_6^{-1/2}$ based on the discussion in Sec.~\ref{sec2} and the fact that $\tau$ at the trapped surface scales with $M^{-1/2}$. Since $t_\mathrm{diff}/t_\mathrm{ff} \propto r^{-1}$, which in turn is proportional to $\tau^2$, we select $\xi=2$ for this study.  

To compute the optical depth $\tau$ numerically, we avoid integrating the opacity along the radial direction as in equation \eqref{eq:tau_est} because of the system's asymmetry. Instead, $\tau$ should be found as the minimum value over any path $\mathcal{C}$ that extends from the boundary of the computational domain to the grid point:
\begin{align}
\tau(r)=- \min \int_\mathcal{C} \rho(r') \kappa(r') dl',
\end{align}
which can be expressed as the Eikonal equation $|\partial_i \tau| = \rho \kappa$.
We implemented the fast sweeping method \citep{Zhao2005} in \texttt{SACRA-2D} to solve for $\tau$; details and tests are provided in Appendix~\ref{A1}.
For simplicity, $\kappa$ is assumed to be constant, $0.35\,\mathrm{g/cm^2}$. 

Viscous hydrodynamics simulation is performed assuming a shear viscous coefficient $\nu_\mathrm{vis}$ with
\begin{equation}
\nu_\mathrm{vis}=\alpha_\mathrm{vis} c_\mathrm{s}^2 \Omega^{-1}, \label{eq27}
\end{equation}
where $\alpha_\mathrm{vis}$ is the so-called alpha parameter~\citep{Shakura1973a} which we assume $O(10^{-2})$, $c_\mathrm{s}$ the sound speed, and $\Omega$ the angular velocity. For a given value of $\Gamma_\mathrm{th}$, $c_\mathrm{s}^2$ is calculated by 
\begin{align}
c_\mathrm{s}^2={1 \over h}{\partial P \over \partial \rho}\Bigg|_{\rm s=const}
%\nonumber \\&
={1 \over h}\left[K\Gamma\rho^{\Gamma-1}+\Gamma_\mathrm{th}(\Gamma_\mathrm{th}-1)(\varepsilon-\varepsilon_0)
\right].
\end{align}
%where $i=1$ or $2$. 
Here we distinguish the general sound speed $c_\mathrm{s}$ from that of an isothermal fluid $a$. 

Using the cylindrical radius and $\nu_\mathrm{vis}$, a viscous timescale is defined by
\begin{equation}
t_\mathrm{vis}:={R^2 \over \nu_\mathrm{vis}}={(R\Omega)^2 \over \alpha_\mathrm{vis} c_\mathrm{s}^2} \Omega^{-1}.
\end{equation}
As this definition shows, viscous effects play an important role for large values of $c_\mathrm{s}/(R\Omega)$ and $\Omega$, i.e., for inner regions. 

Since $c_\mathrm{s}$ is assumed to be $\approx 10^6$\,cm/s for $\varepsilon=\varepsilon_0$, 
the sound speed is much slower than $R\Omega$ near the black hole (at which $R\Omega \sim 10^{10}$\,cm/s) in the absence of shock heating effects. This indicates that viscous effects are significant only when $\varepsilon \gg \varepsilon_0$, which is driven by shock heating. Such conditions occur only in the photon-trapped region where $\Gamma_\mathrm{th}=4/3$. 

\subsection{Setup} \label{sec3.2}

\begin{table}
\caption{Parameters for each model. From left, model name, black hole mass, specific angular momentum parameters $j_0$, and alpha parameter.
The last column shows the presence (Y) or absence (N) of an outflow along the $z$-axis. "S" indicates that the outflow is launched but eventually stalls within the computational region. For model M6.02.05, a strong outflow is launched, but this may be an artifact of the numerical setup; see the text on this. 
\label{table1}}
\begin{tabular}{lcccc}
\hline
Model & $M~(M_\odot)$ & $j_0/M$ %& $n$ 
& $\alpha_\mathrm{vis}$ 
        %& $\Gamma_\mathrm{th}$ 
& Outflow  \\ \hline
M6.30.05   & $10^{6}$   & $30$ &  $0.05$ & S \\ \hline   
M6.30.02   & $10^{6}$   & $30$ &  $0.02$ & S\\ \hline   
M6.30.01   & $10^{6}$   & $30$ &  $0.01$ & S\\ \hline   
M6.20.05   & $10^{6}$   & $20$ &  $0.05$ &  S \\ \hline 
M6.10.05   & $10^{6}$   & $10$ &  $0.05$ &  S \\ \hline 
M6.04.05   & $10^{6}$   & $4$ &  $0.05$ &  Y \\ \hline 
M6.02.05   & $10^{6}$   & $2$ &  $0.05$ &  Y* \\ \hline 
M5.30.05   & $10^{5}$   & $30$ &  $0.05$ & Y \\ \hline
M5.10.05   & $10^{5}$   & $10$ &  $0.05$ & Y \\ \hline 
M6.5.30.05   & $10^{6.5}$   & $30$ & $0.05$ & N \\ \hline 
M6.5.10.05   & $10^{6.5}$   & $10$ & $0.05$ & N \\ \hline 
M6.5.04.05   & $10^{6.5}$   & $4$ & $0.05$ & S \\ \hline 
M7.30.05   & $10^{7}$   & $30$ & $0.05$ & N \\ \hline 
M7.10.05   & $10^{7}$   & $10$ & $0.05$ & N \\ \hline
M7.04.05   & $10^{7}$   & $4$ & $0.05$ & N \\ \hline
\end{tabular}
\end{table}

As summarized in Sec.~\ref{sec2}, this problem involves three regions, assuming that the black hole mass ranges from $10^5M_\odot$ to $10^7M_\odot$. For the outer region where $r \geq r_{\mathrm{ps}}$, the flow is nearly isothermal. In the intermediate region with $r_{\mathrm{trap}} \leq r \leq r_{\mathrm{ps}}$, shock heating occurs, alongside photon diffusion cooling. Inside the innermost region, with $r \leq r_{\mathrm{trap}}$, photons are trapped, making the flow adiabatic. There, shocks heat the matter, radiation pressure dominates, and the adiabatic index is $4/3$. 

The photon-trapped region is larger for less massive black holes, as $r_\mathrm{trap}/r_\mathrm{g} \propto M^{-1}$ (see equation~\eqref{eq15}). This means that shock heating is more efficient at lower black hole masses. To explore how the results depend on black hole mass, we conduct simulations for $M=10^5$, $10^6$, $10^{6.5}$, and $10^7M_\odot$. 

The specific angular momentum values are set to $j_0/M=2$, $4$, $10$, $20$, or $30$. For these values, the centrifugal force is significantly weaker than the gravitational force for $r \gtrsim 10^3r_\mathrm{g}$. Nonetheless, a torus forms in the inner region when $j_0 \gtrsim 4M$, with a typical radius of $j_0^2/M$, and it acts as a heating source through shocks and viscous heating. We observed that for $j_0=2M$, the angular momentum is too small to form a torus without viscosity. However, with viscous angular momentum transport, this is no longer the case (see Sec.~\ref{sec3.3}).  

Numerical simulations are performed for $\alpha_\mathrm{vis}=0$, 0.01, 0.02, and $0.05$. Table~\ref{table1} lists the parameters of the simulations. Each numerical simulation is performed for a timescale longer than $5 \times 10^6t_\mathrm{g} \approx 2.5\times 10^7 M_6$\,s. 

Simulations are performed using the new code {\tt SACRA-2D}, developed recently~\citep{Lam:2025pmz}. For this work, we implemented a (shear) viscous hydrodynamics solver based on the formalism described in~\cite{shibata2017b}.
The detailed implementation of viscous hydrodynamics in {\tt SACRA-2D} is presented in Appendix \ref{A2}.
We employ the two-to-one fixed-mesh-refinement (FMR) structure in the computational domain, which is composed of a hierarchy of nested concentric grids overlaying on top of each other.
The computational domain covers the region of $[0:x_{\rm max}]$ and $[0:z_{\rm max}]$ for $x$ and $z$, respectively, where $x$ denotes the cylindrical coordinate.
It consists of $L$ levels of FMR domains, each of which contains an even number of grids $N$ in the $x$ and $z$ directions, with the grid spacing written as
\begin{align}
\begin{split}
    \Delta x^{(0)}=\Delta z^{(0)} &= x_{\max} / N, \\
    \Delta x^{(l)}=\Delta z^{(l)} &= \Delta x^{(l-1)} / 2, 
\end{split}
\end{align}
for $l=1,2,\cdots,L-1$. Levels $0$ and $(L-1)$ represent the coarsest and finest levels, respectively.
Hydrodynamic variables are set at cell-centered locations. We choose $L=18$, $N=64$, and $x_{\rm max} = z_{\rm max} = 4\times 10^5\,r_{\rm g}$ for this study. Consequently, the computational domain lies well within the sonic point. In this configuration, the grid spacing in the finest region is approximately 0.0477 $r_{\rm g}$. The black hole horizon is located at $r_{\rm H} = 2r_{\rm g}$ in our chosen coordinates (Eddington-Finkelstein). The free-fall timescale at $x_{\rm max}$ and $z_{\rm max}$ is about $2.5\times 10^8 t_{\rm g}$, significantly longer than the total simulation duration. Therefore, we initialize the system with a Bondi solution and do not inject matter from the outer boundary during the evolution.

\subsection{Numerical results}\label{sec3.3}

\begin{figure}
    \centering
    \includegraphics[width=\columnwidth]{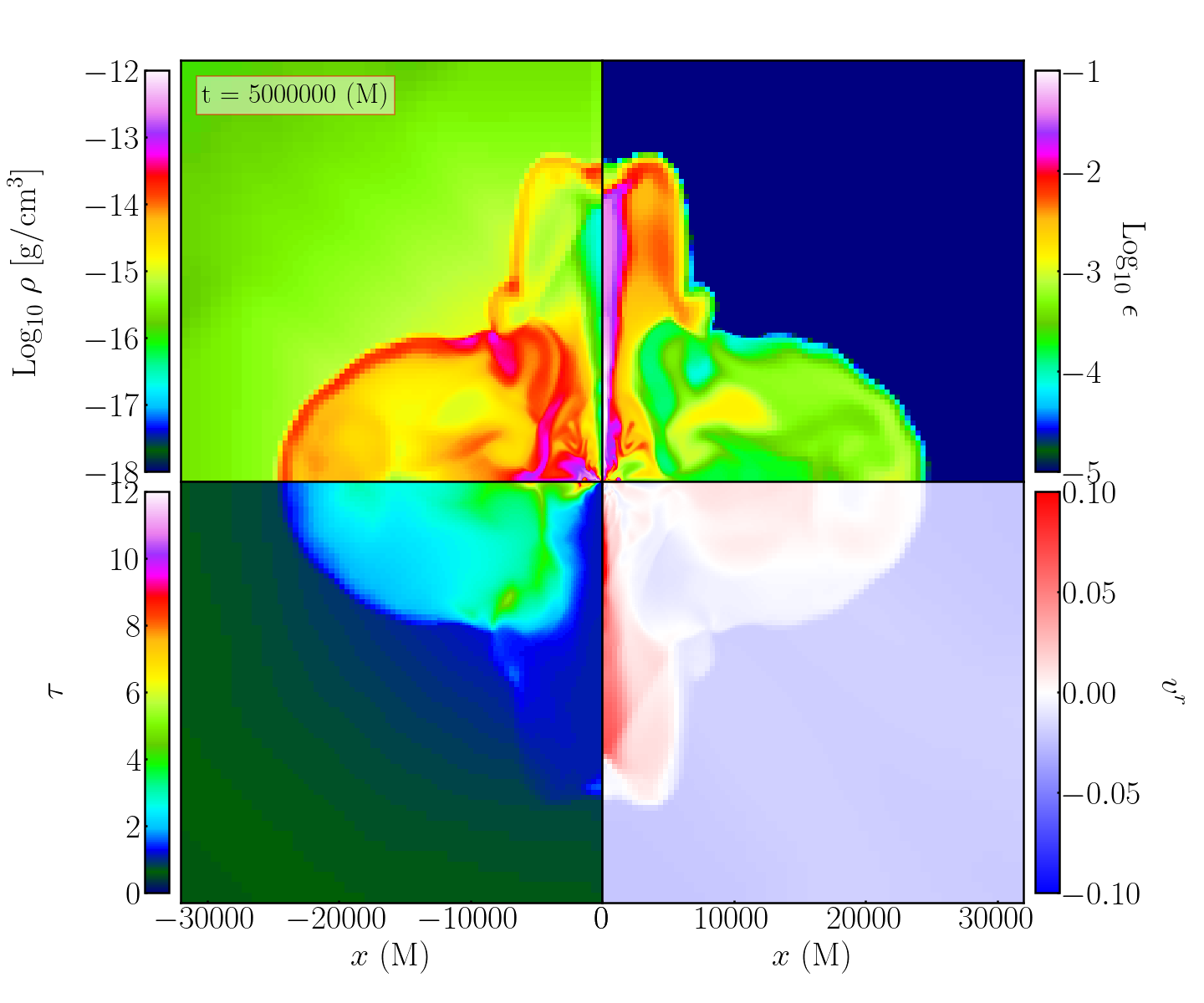}
    \caption{Snapshot of the rest-mass density (top left), specific internal energy in units of $c^2$ (top right), opacity (bottom left), and radial velocity $v^r := \gamma^{ri} u_i /w$ in units of $c$ (bottom right) for model M6.30.05 at $t = 5 \times 10^6 t_\mathrm{g}$. Note that the length unit is $r_\mathrm{g} \approx 1.48 \times 10^{11}$\,cm for $M=10^6M_\odot$. Animations for this model are available at 
    \url{https://www2.yukawa.kyoto-u.ac.jp/~masaru.shibata/M6.30.1.05_lv3_N64.mp4} (large scale) and 
    \url{https://www2.yukawa.kyoto-u.ac.jp/~masaru.shibata/M6.30.1.05_lv5_N64.mp4} (smaller scale).
    }
    \label{fig_M6.30}
\end{figure}

\begin{figure}
    \centering
    \includegraphics[width=\columnwidth]{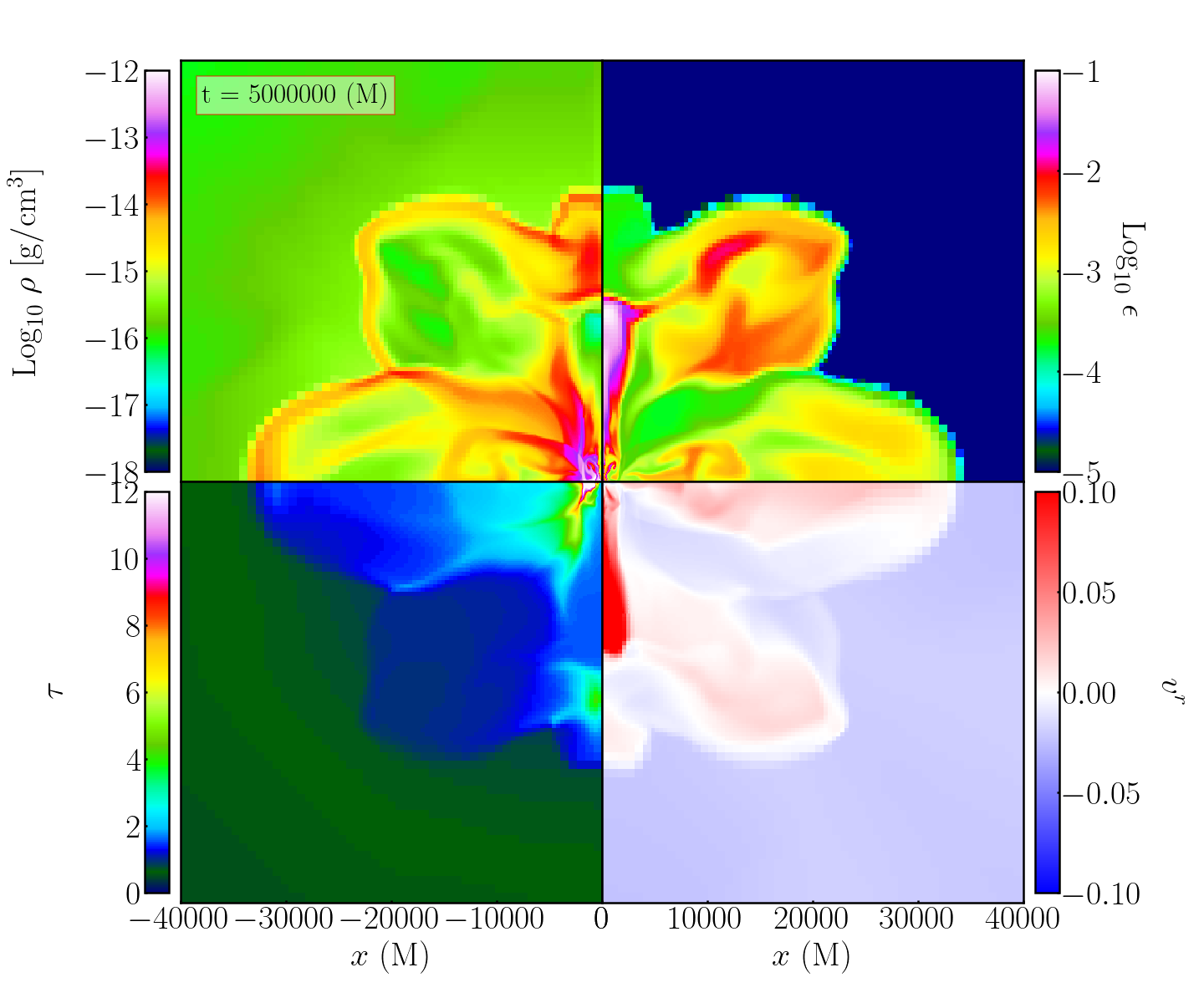}
    \caption{The same as Fig.~\ref{fig_M6.30} but for model M6.10.05. Animations for this model are available at 
    \url{https://www2.yukawa.kyoto-u.ac.jp/~masaru.shibata/M6.10.1.05_lv3_N64.mp4} (large scale) and 
    \url{https://www2.yukawa.kyoto-u.ac.jp/~masaru.shibata/M6.10.1.05_lv5_N64.mp4} (smaller scale). 
    }
    \label{fig_M6.10}
\end{figure}

Using model M6.30.05 as a fiducial case, we first outline the system's evolution (see Fig.~\ref{fig_M6.30} and corresponding animations: cf. the figure caption). 
Soon after the simulation begins, shocks emerge from centrifugal bounce and the collision between infalling matter and a torus formed near the black hole. Shocks propagate outward through the photon-trapped region, where $\Gamma_\mathrm{th}=4/3$, but eventually stall outside this region, where $\Gamma_\mathrm{th}$ drops to about 1.01. After the initial phase, matter accumulates in the photon-trapped region, causing the torus around the black hole to grow and a hot envelope to form around the torus. 

Since the specific angular momentum of the matter infalling around the symmetric axis ($z$-axis) is small, mass accretion onto the black hole also proceeds. However, direct mass accretion onto the black hole is minor for the models with $j_0 \geq 4M$ because it is allowed only for fluid elements with specific angular momentum smaller than $j_\mathrm{crit} \lesssim 4M < j_0$. Consequently, only a fraction of the infalling matter with
\begin{equation}
\sim 1-\sqrt{1-\left({j_\mathrm{crit} \over j_0}\right)^{1/n}}
\end{equation}
could directly fall onto the black hole. Actually, after the formation of a torus, shocked matter is launched toward the polar region, which prevents matter from infalling toward the black hole, and therefore, the mass accretion is further suppressed. 

After the growth of the torus, an outflow is launched along the $z$-axis. The typical radial velocity of the outflow is $\lesssim 0.1c$ (though this could change if magnetohydrodynamical effects were included). The outflow initially penetrates the infalling envelope. However, once the outflow head reaches a low-density region where $\Gamma_\mathrm{th}$ approaches $1.01$, the outflow stalls, and a cocoon is formed toward the direction perpendicular to the $z$-axis. After this stalling, gas accumulates steadily around the black hole, developing a dense gas structure. The dense envelope halts infalling matter at large radii, reducing the shock-heating efficiency in the inner region. Thus, the Eddington mass accretion rate, defined by $L_\mathrm{Edd}/(\eta_\mathrm{Edd}c^2)$ with $\eta_\mathrm{Edd}$ being the heating efficiency, effectively increases due to the decrease of $\eta_\mathrm{Edd}$ from $0.1$ to $\ll 0.1$. This also reduces the outflow power in the late phase. In this model, the heat generated by viscosity drives the outflow primarily along the $z$-axis, with a minor fraction driving convective motion in the torus. 

For model M6.10.05, the torus can be more compact due to the smaller specific angular momentum, and hence the viscous heating efficiency near the black hole can be higher (see Fig.~\ref{fig_M6.10}). 
This results in a stronger outflow along the $z$-axis than for model M6.30.05, and the outflow head extends farther (compare Figs.~\ref{fig_M6.30} and \ref{fig_M6.10}). The outflow is even stronger and extends farther for model M6.04.05~\footnote{See the animation for model M6.04.05 at 
\url{https://www2.yukawa.kyoto-u.ac.jp/~masaru.shibata/M6.04.1.05_lv3_N64.mp4}.}. 
However, the outflow eventually stalls in the region with $\tau \lesssim 1$, and the resulting configuration after stalling is qualitatively the same as for model M6.30.05; the eventual configuration is universally composed of an expanding torus near the equatorial plane and a stalled outflow along the $z$-axis for the models with $M=10^6M_\odot$. 

For model M6.02.05, a strong outflow is also launched in a late stage, while initially most of the inflowing matter falls into the black hole. The outflow results from late-time formation of a torus due to the long-term viscous transport of angular momentum within the inflowing matter; matter coming from distant regions, especially near the equatorial plane (since $\nu_\mathrm{vis} \propto R^{3/2}$), can acquire significant angular momentum due to viscous effects in the current setup, leading to the formation of a torus and an outflow. However, this may be a computational artifact, as strong viscous effects are expected primarily in the torus or disc, not in the infalling material. 

\begin{figure}
    \centering
    \includegraphics[width=\columnwidth]{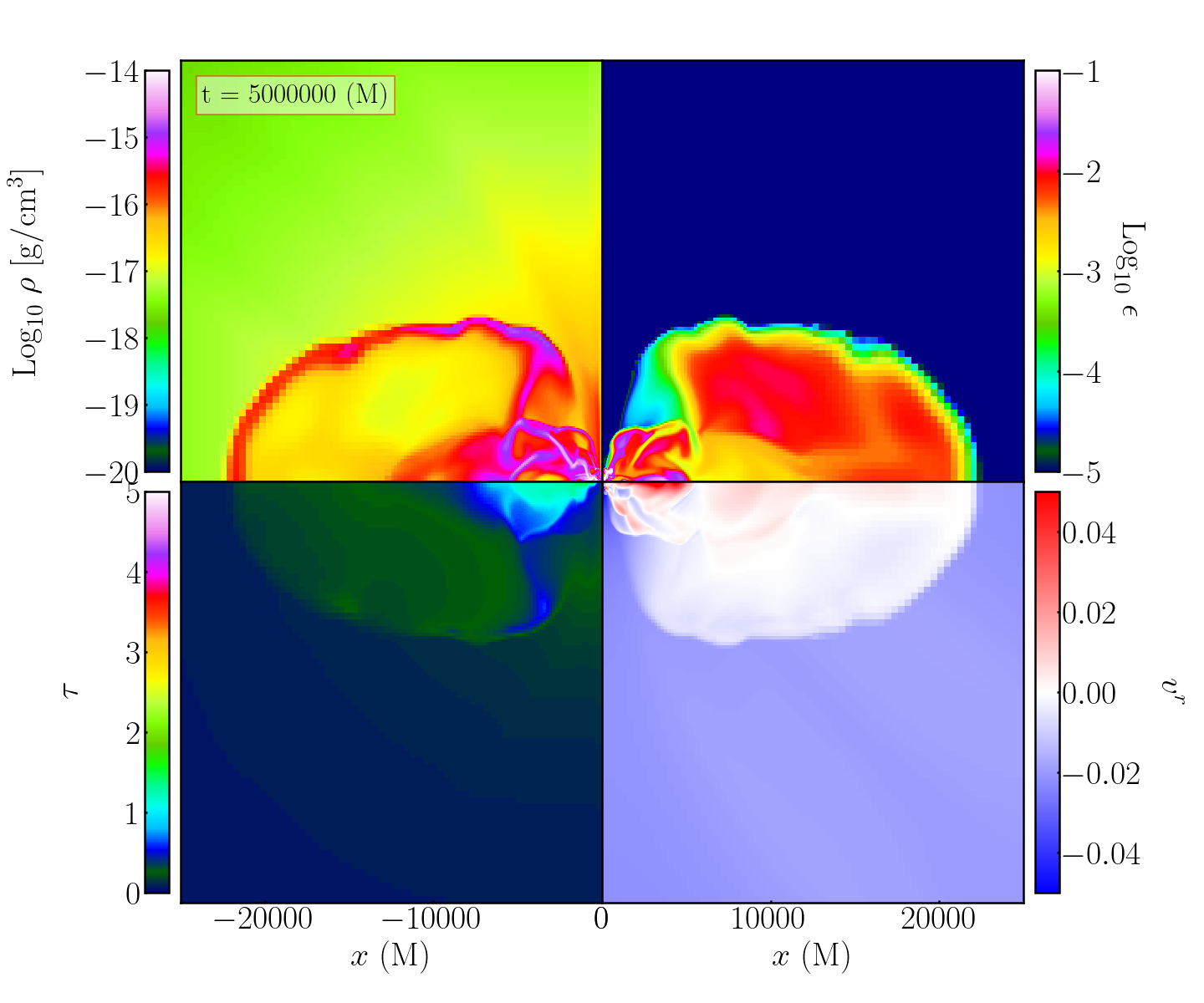}
    \caption{The same as Fig.~\ref{fig_M6.30} but for model M7.30.05 at $t = 5 \times 10^6 t_\mathrm{g}$. Note that the length unit is $r_\mathrm{g} \approx 1.48 \times 10^{12}$\,cm for $M=10^7M_\odot$. 
    Animations for this model are available at 
    \url{https://www2.yukawa.kyoto-u.ac.jp/~masaru.shibata/M7.30.1.05_lv3_N64.mp4} (large scale) and 
    \url{https://www2.yukawa.kyoto-u.ac.jp/~masaru.shibata/M7.30.1.05_lv5_N64.mp4} (smaller scale).
    }
    \label{fig_M7.30}
\end{figure}

For larger black hole masses ($M=10^{6.5}M_\odot$ and $10^7M_\odot$), where the trapped region is narrower, the launch of the outflow along the $z$-axis is suppressed. In particular, for $M=10^7M_\odot$, no outflow is launched during the simulation time $t \gtrsim 5 \times 10^6t_\mathrm{g}$: see Fig.~\ref{fig_M7.30} for model M7.30.05. Indeed, for this model, the optically thick region with $\tau \geq 1$ is so narrow that the energy generation efficiency is too low to launch jets. Nevertheless, thermal energy is generated by viscous and shock heatings, and transported by convection from the inner to the outer region of the torus, leading to the formation of a hot convective torus (see the plot for $\varepsilon$ in Fig.~\ref{fig_M7.30}).

In models where the black hole mass $M \geq 10^{6.5}M_\odot$, the hot torus surrounding the black hole expands regardless of the values of $j_0$. This suggests that once the black hole mass exceeds approximately $10^{6.5}M_\odot$, the torus undergoes a quasi-steady evolution without outflow activity, reaching a quasi-star phase~\citep{Begelman:2007je}. For these high-mass black hole models, the viscosity-generated heat mainly heats the torus matter via convective motion, as there is no outflow along the $z$-axis~\citep{Coughlin_2024}. Possible outcome after the further growth of the envelope will be discussed in Sec.~\ref{sec4}.

In contrast to high black hole mass models, the $M=10^5M_\odot$ case exhibits an energetic outflow launched along the $z$-axis after the torus is formed~
\footnote{See the animation for model M5.30.05 at 
\url{https://www2.yukawa.kyoto-u.ac.jp/~masaru.shibata/M6.04.1.05_lv3_N64.mp4}.}.
This outflow continuously penetrates the infalling matter without stalling and eventually reaches the outer boundary of the computational domain at $4 \times 10^5 r_\mathrm{g}$. However, in regions far from the center where the rest-mass density, ionization, and opacity are sufficiently low, the outflow will eventually stall due to inefficient heating. It is important to note that the outflow is collimated, resulting in negligible mass loss. For this low-mass model, where $\dot M_\mathrm{b} < \dot M_\mathrm{Edd} \ll \dot M$ with $\dot M_\mathrm{b}$ denoting the rest-mass accretion rate onto the black hole, the envelope mass increases linearly with $\dot M t$ approximately.

%%%
\begin{figure}
    \centering
    \includegraphics[width=\columnwidth]{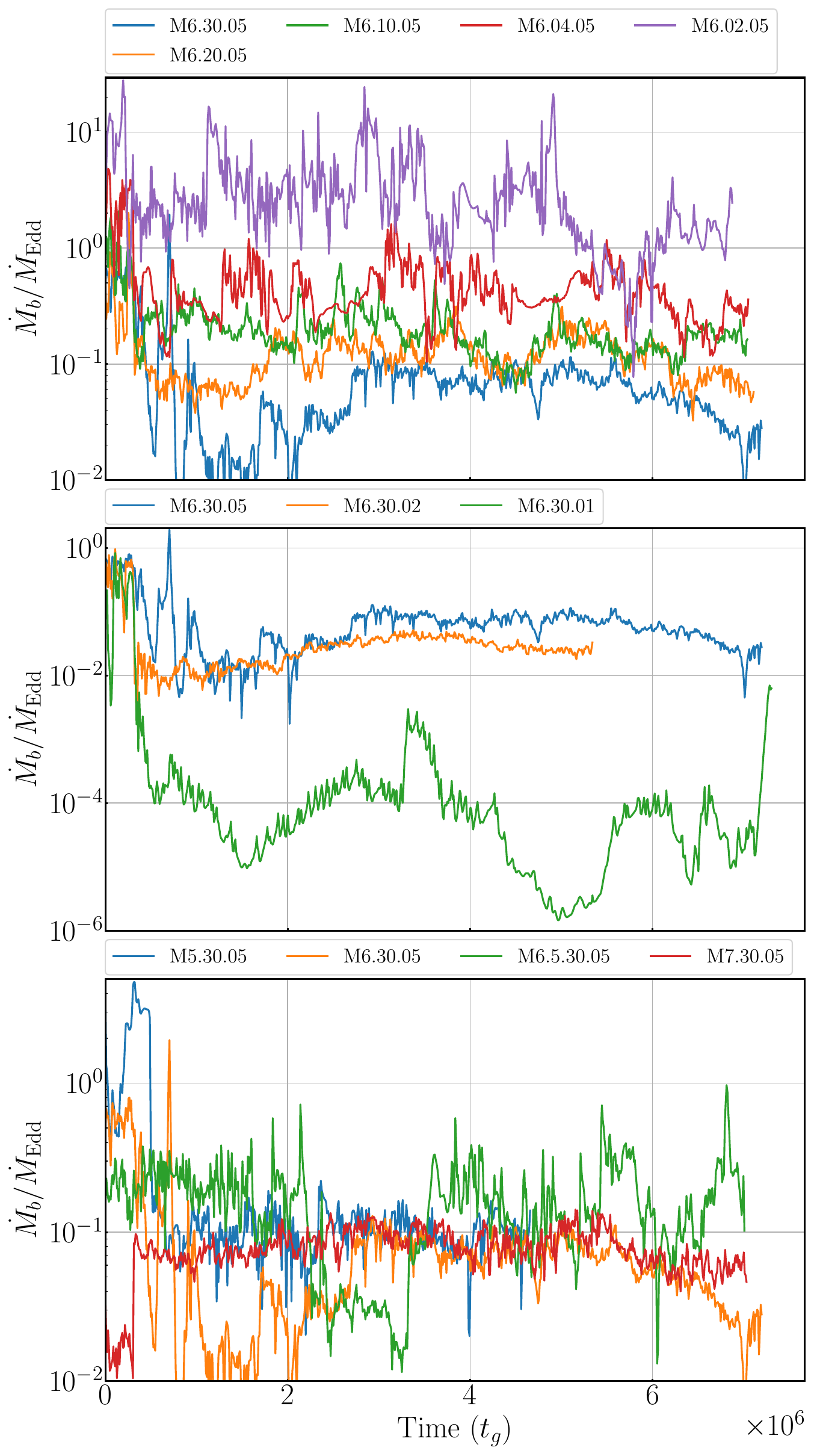}
    \caption{The rest-mass accretion rate $\dot M_\mathrm{b}$ for $(M, \alpha_{\rm vis})=(10^6M_\odot, 0.05)$ (top),
    for $(M, j_0)=(10^6M_\odot, 30 M)$ (middle),
    and $(j_0, \alpha_{\rm vis})=(30 M, 0.05)$ (bottom) in units of $\dot M_{\rm Edd}$. 
    }
    \label{fig_acc}
\end{figure}

Figure~\ref{fig_acc} shows the evolution of the mass accretion rate onto the black hole in units of $\dot M_\mathrm{Edd}$ for models with $(M, \alpha_{\rm vis})=(10^6M_\odot, 0.05)$ with different values of $j_0$ in the top panel, for models with $(M, j_0)=(10^6M_\odot, 30 M)$ with different values of $\alpha_\mathrm{vis}$ in the middle panel, and for models with $(j_0, \alpha_{\rm vis})=(30 M, 0.05)$ with different black hole masses in the bottom panel.
The figure shows that for $j_0 \geq 4M$ and $\alpha_\mathrm{vis}=0.05$, the value of $\dot M_\mathrm{b}/\dot M_\mathrm{Edd}$ is of order 0.1--1. 
Recalling equation~\eqref{eq19}, we find that the growth timescale of the BHs, $t_\mathrm{Edd}$, is of order $10^8$\,yrs in the present setting, irrespective of the black-hole mass. The upper panel of Fig.~\ref{fig_acc} shows that for smaller values of $j_0$, the accretion rate onto the black hole is naturally larger. Especially for $j_0=2M$, a substantial fraction of the matter directly falls into the black hole.
The middle panel of Fig.~\ref{fig_acc} shows that a small viscosity $\alpha_{\rm vis}=0.01$ suppresses the accretion rate onto the black hole because the longer viscous timescale for angular momentum transport delays matter infall.
Nonetheless, the black hole mass accretion rate in units of $\dot M_\mathrm{Edd}$ depends only weakly on the black hole mass for the reasonable values of $\alpha_\mathrm{vis} \geq 0.02$.

For low black-hole masses like $M=10^5M_\odot$ and $10^6M_\odot$, the injected mass accretion rate at the sonic point $\dot M$ is much larger than $\dot M_\mathrm{Edd}$. This implies that only a minor fraction of $\dot M$ falls into the black hole for the low black-hole mass models for $j_0 \geq 4M$, i.e., $\dot M_\mathrm{b} \ll \dot M$. The implication of this result will be discussed in Sec.~\ref{sec4}. 

\begin{figure}
    \centering
    \includegraphics[width=\columnwidth]{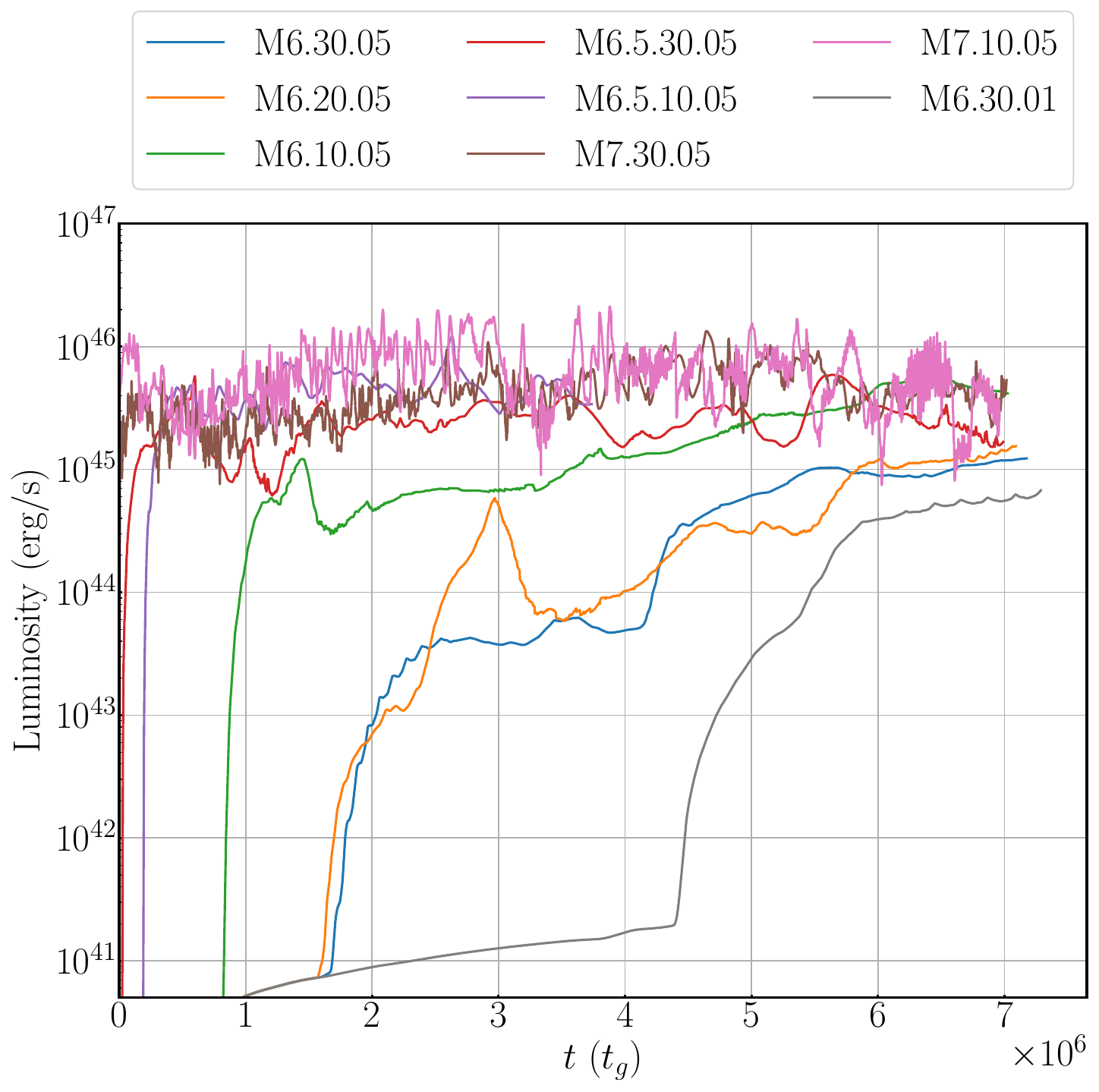}
    \caption{The luminosity curve for selected models with $M=10^6$, $10^{6.5}$, and $10^7M_\odot$.
    }
    \label{fig_lumi}
\end{figure}

We broadly estimate the luminosity from the black hole envelope. For this purpose, we assume that the remnant is quasi-steady with the energy injection from the inner region. Under this assumption, we determine the surface of $\tau=1$, evaluate the temperature $T$ assuming $a_\mathrm{r}T^4=\rho(\varepsilon-\varepsilon_0)$ for $\tau \geq 1$, and estimate the luminosity by
\begin{equation}
L=\oint_{\tau=1} \sigma T^4 \hat{r} \cdot d\hat S
={c \over 4}\oint_{\tau=1}  \rho(\varepsilon-\varepsilon_0) \hat{r} \cdot d\hat S,
\end{equation}
where $d\hat S$ and $\hat r$ are the area element vector and unit normal vector at the surface of $\tau=1$, respectively. For models with low black hole masses, this method may not be appropriate because the $\tau=1$ surface is expanding at an appreciable velocity, and thus the photon-emission surface should be determined by the condition $\tau = c/V$, where $V$ denotes the expansion velocity. Thus, we here pay attention only to the model with $M \geq 10^6M_\odot$ for which the envelope with $\tau \geq 1$ is considered to be in a quasi-steady state at late times. 
We also note that for model M7.30.05, $\tau$ could remain below 1 all the way to black holes at the polar region. We simply set $T=0$ for those values of $\theta$ when integrating luminosity.

Figure~\ref{fig_lumi} shows the light curve obtained in our simple prescription for models with $M=10^6$, $10^{6.5}$, and $10^7M_\odot$. This shows that for $M=10^{6.5}$ and $10^7M_\odot$ the luminosity relaxes to a value of the order of $L_\mathrm{Edd}$ (see equation~\eqref{eq17}). This is reasonable because the high-density part inside the $\tau=1$ surface is radiation-dominant and evolves quasi-steadily. This suggests that the black hole-low-mass envelope systems in the growth phase would be observed as a quasi-steady system with the Eddington luminosity defined by the black hole mass even when the envelope mass is much smaller than the black hole mass (note that the envelope mass in these models is smaller than $M_\odot$). For $M=10^6M_\odot$, the luminosity is lower than $L_\mathrm{Edd}$ for early stages with $t\lesssim 3 \times 10^6t_\mathrm{g}$, and then, becomes higher than $L_\mathrm{Edd}$. This super-Eddington behavior is likely associated with the outflow launch and resulting increase of the shocked material. Thus, for low-mass black holes, the luminosity may a bit exceed the Eddington luminosity for growing SMBHs. 

As found from (e.g., for $M=10^7M_\odot$), 
\begin{align}
T_\mathrm{ave} =\left[{L \over 4\pi \sigma r_{\tau=1}^2}\right]^{1/4}
&\approx  9.8\times 10^3 \,\mathrm{K}\left({L \over L_\mathrm{Edd}}\right)^{1/4}
\nonumber \\&~~\times 
\left({r_{\tau=1} \over 10^4 r_\mathrm{g}}\right)^{-1/2} 
\left({M \over 10^7M_\odot}\right)^{-1/2}, 
\end{align}
where $r_{\tau=1}$ denotes the typical radius of $\tau=1$ surface, the average temperature is $T_\mathrm{ave}\sim 10^4$\,K, higher than the temperature of the infalling nearly isothermal gas with $T\approx 7000$\,K. This is because of the presence of shock heating and heating by convection of a hot gas supplied from the inner region. As indicated in Figs.~\ref{fig_M6.30}--\ref{fig_M7.30}, the estimated temperature at $\tau=1$ surfaces is highly non-uniform, and locally, it can be several times $10^4$ \,K in the presence of an efficient heating.

\section{Discussions}\label{sec4}

This section examines the likely growth trajectory of SMBHs based on the numerical results in this paper.
Our results indicate that, regardless of black hole mass, the mass accretion rate, $\dot M_\mathrm{b}$, is of order 10\% of $\dot M_\mathrm{Edd}$ for the reasonable choice of $\alpha_\mathrm{vis}$ Consequently, the black hole's growth timescale, $t_\mathrm{BH}$, is of order $10^8$\,yrs irrespective of the black hole mass (see equation~\eqref{eq19}). For low-mass black holes, the rate at which the envelope accumulates mass surpasses $\dot M_\mathrm{b}$ because the ratio $\dot M/\dot M_\mathrm{Edd} \approx 9.4M_6^{-1}$ in our model. This suggests that, for black holes with $M \lesssim 10^6 M_\odot$, the envelope mass, $\simeq \dot M t$, could exceed the black hole mass, which grows as $M_0 \exp(\eta t/t_\mathrm{Edd})$, where $M_0$ is the initial mass and $\eta=\dot M_\mathrm{b}/\dot M_\mathrm{Edd}=O(0.1)$, during growth. Such a quasi-star scenario may be a natural consequence~\citep{Begelman:2007je} if the mass input rate stays near $\sim \dot M \approx 1.5\times 10^{25}$ g/s (see equation~\eqref{eq16}) for over $10^8$ years and the envelope mass accumulation continues steadily. For an initial mass $M_0=10^m M_\odot$ with $m=5$--6, the evolution over a time approximately $t_\mathrm{Edd}/\eta \sim 10^8$ years would produce a black hole of mass $\sim e^{1} \times 10^m M_\odot$ and an envelope mass $M_\mathrm{env}\approx 0.94/\eta \times 10^7M_\odot$. Hence, in the later stages, the envelope's self-gravity could become critical in determining its density and velocity profiles. 
Note, however, that the evolution process of the system may be different from this naive picture when the envelope mass becomes comparable to the black hole mass (see below).

By contrast, if the mass injection rate varies with time and the injection timescale is shorter than $\sim 10^8 (M/10^7M_\odot)$\,yrs, the envelope mass does not exceed the black hole mass. In such a case, a black hole-low-mass envelope state may be the typical consequence. However, in this scenario, the intermittent timescale must be longer than the Eddington timescale, so the rapid growth of the black hole cannot occur. Assuming the rapid growth of black holes in the early universe, the quasi-star systems may be more common than the black hole-low-mass envelope systems for low-mass black holes with $M\lesssim 10^6M_\odot$. 

Although a jet can penetrate the envelope in the early phase of the evolution for which $M_\mathrm{BH} > M_\mathrm{env}$, this may be prohibited in the presence of dense matter and resulting self-gravity in the later evolution stage. This point should be studied in future work. 

%%%%%
In the present model, the temperature at the photosphere with $\tau=1$ is $\gtrsim 10^4$\,K and the resulting luminosity is broadly Eddington. Assuming that the radius of the photosphere should be approximately constant for a given luminosity $\sim L_\mathrm{Edd}$ and temperature, with the increase of the envelope's mass, the average density inside the envelope should increase. Then, the viscous heating rate appears to increase, but the luminosity is unlikely to exceed the Eddington luminosity significantly in the absence of strong outflows. Then, convection should modify the matter profile to reduce the viscous heating rate, and as a result, the matter density (or mass accretion rate) in the vicinity of the black hole is likely to be reduced (i.e., the mass accretion rate near the black hole is controlled to be low). Such a state is predicted in convection-dominated accretion flow (CDAF)~\citep{2000ApJ...539..798N, 2000ApJ...539..809Q, 2000ApJS..130..463I, 2002ApJ...565.1101A}, and therefore, it is likely to be the expected resulting state in the inner region. 

\begin{figure}
    \centering
    \includegraphics[width=\columnwidth]{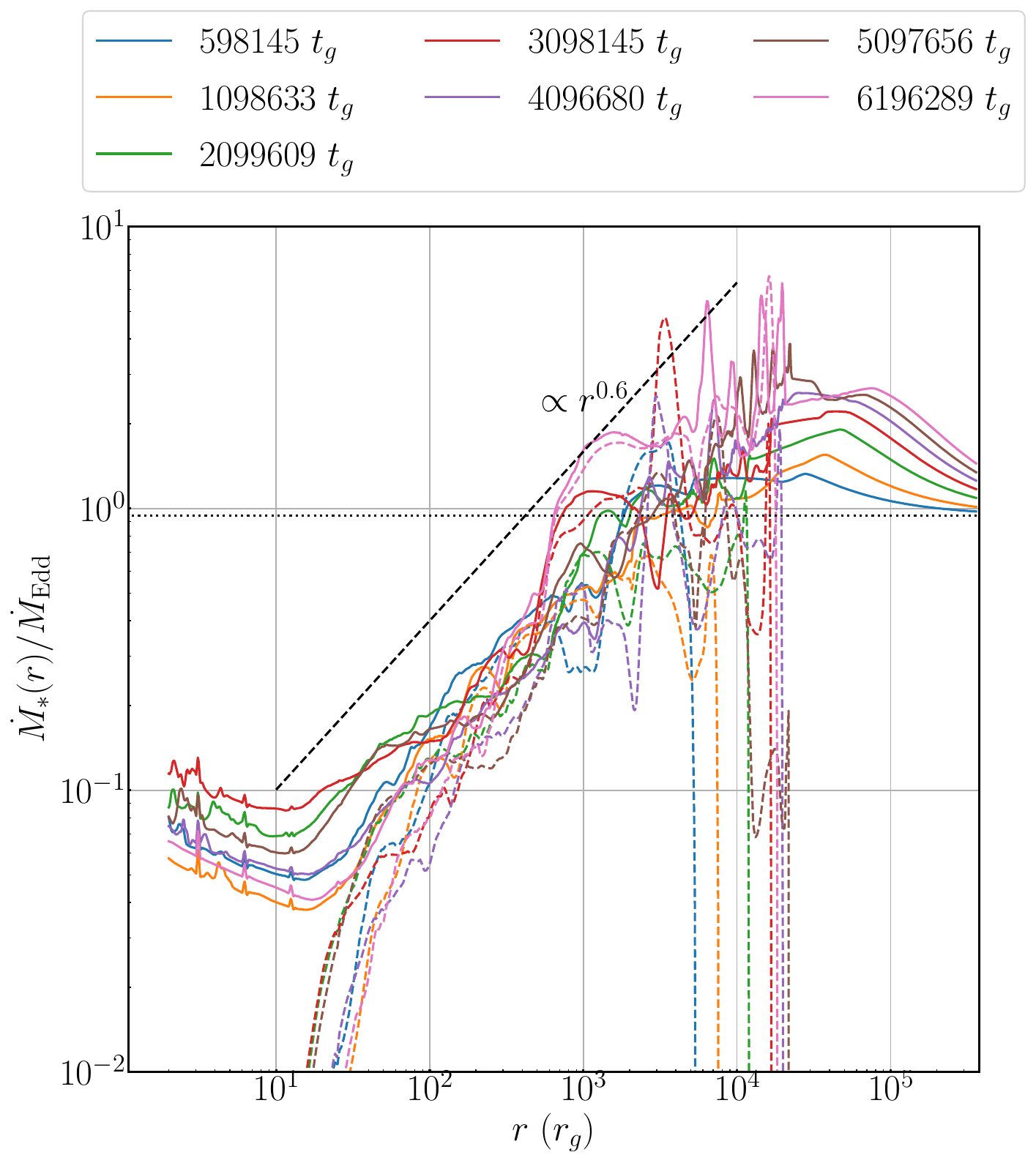}
    \caption{Inflow and outflow mass accretion rates (solid and dashed curves, respectively) in units of $\dot M_\mathrm{Edd}$ as functions of the radius at selected time slices for model M7.30.05.
    }
    \label{fig_dotm}
\end{figure}

Indeed, such a CDAF state is seen in our results.
We analyze the local mass accretion rate for the models with $M=10^7M_\odot$, for which the system relaxes to a quasi-steady state in the late stage. The analysis is carried out in the same manner as in \citet{Shibata:2025ycs} (see equations (18)--(20) of this paper). Figure~\ref{fig_dotm} displays the ingoing and outgoing mass inflow rates (solid and dashed curves, respectively) as functions of the radius at selected time slices for model M7.30.05. The dotted horizontal line denotes the injection rate, $\dot M$. It is found for $r \lesssim 100 r_\mathrm{g}$, the inflow is dominant with the mass accretion of $\sim 0.1\dot M$. 
This indicates that the density is low in such a region due to the presence of convection in the outer region. For $10^2 \lesssim r/r_\mathrm{g} \lesssim 10^4$, the inflow and outflow rates are comparable, and the inflow rate is broadly proportional to $r^\beta$ with $\beta\sim 0.6$. This relation is in good agreement with that found in \citet{Shibata:2025ycs}, indicating the establishment of a convection-dominated state. The fact that the mass inflow and outflow rates are comparable in this region, which is roughly where the photosphere should be located (see Eq.~\eqref{eq:r_ps}), is in line with the analysis of~\cite{2026arXiv260118864S} where a similar conclusion is reached based on the degree of symmetry in the exponential wings of lines in LRD spectra. We note that for $r \gtrsim 10^4r_\mathrm{g}$ the inflow rate is increasing. The reason for this is that the matter is accumulating outside the shock surface with $\tau \sim 1$ (cf. Fig.~\ref{fig_M7.30}). We also note that for lower mass cases with $M \leq 10^6M_\odot$, in which $\dot M \gg \dot M_\mathrm{b}$ the power, $\beta$, is naturally larger. However, for these models, the system does not relax to a quasi-steady state in the simulation time, so the final power is not very clear.

As mentioned in the first paragraph of this section, the envelope could be self-gravitating after a substantial accretion proceeds. In such a situation, the structure of the envelope may be significantly different from that we found in this paper due to the effects that were not taken into account. First, the self-gravity effect will enhance non-axisymmetric deformation of the torus and envelope (see, e.g., \citealt{Shibata:2021sau}), and hence, the mass accretion rate may be significantly increased by the enhanced gravitational torque. It is also not clear whether the viscous heating rate is always comparable to $L_\mathrm{Edd}$ even for high envelope masses because the order of the magnitude for the viscous heating rate is estimated as $\dot E_\mathrm{vis}\sim \alpha_\mathrm{vis} \beta_\Omega^2 M_\mathrm{env} \bar c_\mathrm{s}^2 \bar \Omega$ (see Appendix \ref{A4}), and thus, 
\begin{eqnarray}
\dot E_\mathrm{vis} &\sim& 9 \times 10^{43}\,\mathrm{erg/s}\,
\left({\alpha_\mathrm{vis} \over 0.05}\right)
\left({\beta_\Omega \over 1.5}\right)^2 
\left({M_\mathrm{env} \over M}\right)\nonumber \\
&& ~~~~~~~~~~~~~~\times
\left({\bar c_\mathrm{s} \over 10^6\,\mathrm{cm/s}}\right)^2
\left({G c^{-4}\bar \Omega M \over 10^{-6}}\right), \label{eq34}
\end{eqnarray}
where $\beta_\Omega$ is the typical value of $|\partial \ln\Omega / \partial \ln R|$, and $\bar c_\mathrm{s}$ and $\bar \Omega$ denote typical sound velocity and angular velocity in the envelope, respectively. 

Equation~\eqref{eq34} indicates that for high mass ratios of $M_\mathrm{env}/M \gtrsim 1$, the viscous heating rate can exceed $L_\mathrm{Edd}$ even if the envelope matter concentrates near the photosphere with $\tau=1$ for $M \lesssim 10^6M_\odot$. Note that $\bar c_\mathrm{s}$ should be larger than $10^6$\,cm/s and the photosphere is approximately determined by $\sqrt{L/(4\pi\sigma T^4)}$ with $T\lesssim 10^4$\,K, and thus, the maximum radius is estimated by $\sqrt{L_\mathrm{Edd}/(4\pi \sigma T^4)} \approx 6.2 \times 10^4 r_\mathrm{g} M_6^{-1/2}\kappa_{0.35}^{1/2}T_{7000}^{-2}$.
If the viscous heating rate exceeds $L_\mathrm{Edd}$ for the high envelope masses, the outflow has to be driven, and further growth of the envelope should be prohibited (see Appendix \ref{A4} for an analysis). 
We plan to clarify these possibilities in future work.

The photospheric temperature of the observed LRDs is typically $\sim 5000$\,K~\citep{2025arXiv251121820D}. In our results, the average temperature at $\tau=1$ is higher than this value. Such high temperature results from shock heating between the growing envelope and infalling matter of assumed temperature $\sim 7000$\,K and from convective motion of a hot gas coming from the inner region. This suggests that the LRDs might not be observed in an active accretion phase. 

\section{Summary}\label{sec5}

To study the formation process of black hole-envelope systems, we performed a long-term viscous hydrodynamics simulation for quasi-spherical accretion onto black holes in general relativity varying the black hole mass and specific angular momentum of the infalling matter while fixing the mass accretion rate as $\dot M=a^3/G \approx 1.5 \times 10^{25}$\,g/s which is comparable to the Eddington mass accretion rate for $10^7M_\odot$ black holes. We found that irrespective of the specific angular momentum of the infalling matter, an outflow is launched along the symmetric axis for $M \leq 10^6M_\odot$ for which the mass accretion is super-Eddington. Especially for $M=10^5M_\odot$, the outflow reaches the outer boundary of the computation domain at $4\times 10^5r_\mathrm{g}$. For $M=10^6M_\odot$, the outflow is eventually stalled at $r=3$--$4\times 10^4r_\mathrm{g}$ by the ram pressure of the infalling matter and due to low-shock heating efficiency, leading to a quasi-steady envelope evolution stage. On the other hand, for $M \geq 10^{6.5}M_\odot$, for which the mass accretion is only slightly super-Eddington, any outflow is not driven, and a convective envelope simply grows. This indicates that for the slightly super-Eddington mass accretion case, a BH-envelope structure simply grows with time with no appreciable outflow. 

We also found that irrespective of the model parameters (black hole mass and specific angular momentum ($j_0 \geq 4M$) of the infalling matter), the mass accretion rate onto the black hole $\dot M_\mathrm{b}$ is of order 10\% of the Eddington mass accretion rate for the reasonable values of $\alpha_\mathrm{vis}$. This implies that for low-mass black holes with $M\lesssim 10^6M_\odot$, the growth rate of the envelope is much larger than $\dot M_\mathrm{b}$, and the envelope would be developed much more quickly than the black hole. If the initial black hole mass is smaller than $\sim 10^6M_\odot$, the resulting outcome as a result of a $10^8$\,yrs-long mass accretion may be a black hole surrounded by an envelope with mass comparable to the black hole mass, i.e., a quasi-star~\citep{Begelman:2007je}. 

The subsequent evolution process after the envelop mass becomes comparable to the black hole mass is not very clear. For the high envelope mass, the viscous heating rate $\dot E_\mathrm{vis}$ is likely to be enhanced and can exceed the Eddington luminosity of the system for any matter configuration of the envelope. If this happens, further mass accretion should be prohibited (cf. a discussion in Appendix~\ref{A4}). 
This hypothesis should be examined by general relativistic simulations in the future.

In this work, we performed viscous hydrodynamics simulations. In reality, viscosity is effectively induced by magnetohydrodynamic processes such as the magnetorotational instability and the resulting dynamo~\citep{Balbus:1998ja}, and for a more realistic study, we need to perform a magnetohydrodynamics simulation in three spatial dimensions, although such a simulation is computationally expensive. In magnetohydrodynamics, a jet along the $z$-axis may be stronger because of the presence of the Blandford-Znajek mechanism~(\citealt{Blandford1977}; in the presence of black hole spin) and mass ejection processes associated with a global magnetic field, such as the magneto-centrifugal effect~\citep{Blandford1982jun}, may enhance the mass ejection efficiency. We plan to perform a magnetohydrodynamics simulation in the subsequent step. 
Moreover, in this work, we employed a simplified equation of state assuming an idealized cooling process for the optical thin region. In reality, cooling is determined by radiation. Thus, incorporating radiation effects is also an important next step.

\section*{Acknowledgements}
The authors thank Kohei Inayoshi for useful discussions. 
Numerical computation was performed on the clusters, Sakura and Momiji, at the Max Planck Computing and Data Facility. The authors benefited from discussions during the Yukawa Institute for Theoretical Physics (YITP) workshop YITP-T-25-02, ``Multi-Messenger Astrophysics in the Dynamic Universe''. 
This work was in part supported by Grant-in-Aid for Scientific Research (grant No.~23H04900 and 23H01169) of Japanese MEXT/JSPS and JST FOREST Program (JPMJFR2136).
A.T.L.L. acknowledges support by NASA under award No. 80NSSC25K7213.

\bibliographystyle{mnras}
\bibliography{reference}

\appendix

\section{Fast sweeping method for Eikonal equation} \label{A1}
To solve the Eikonal equation for the optical depth $\tau$, we follow \cite{Zhao2005} to implement the fast sweeping method in \texttt{SACRA-2D}, which is briefly outlined here.
The equation is discretized by Godunov's upwind difference scheme at the grid point $(i,j)$ to
\begin{align}
    \left[ \left(\tau_{i,j} - \min(\tau_{i\pm 1,j}) \right)^+ \right]^2 +
    \left[ \left(\tau_{i,j} - \min(\tau_{i,j\pm 1}) \right)^+ \right]^2
    = f_{i,j}^2 h^2,
\end{align}
where $f_{i,j}=\kappa \rho_{i,j}$ is the source term, $h=\Delta x = \Delta z$ is the grid size, and $(x)^+:= \max(x, 0)$.
Initially, we set $\tau$ at the boundary of the computational domain using equation~\ref{eq:tau_est}. For all interior grid points, we assign large initial values (e.g., $10^{99}$) to $\tau_{i,j}$, which will be updated in subsequent iterations.
In each iteration, we cycle through the domain using four alternate grid point orders for $i,j=1,\dots,N$:
(i) $i=1$ to $N$ and $j=1$ to $N$;
(ii) $i=1$ to $N$ and $j=N$ to $1$;
(iii) $i=N$ to $1$ and $j=N$ to $1$;
(iv) $i=N$ to $1$ and $j=1$ to $N$, to update $\tau_{i,j}$ as
\begin{align}
    \tau_{i,j}^{\rm new} =
    \begin{cases}
        \min(\tau_1, \tau_2) + f_{i,j} h,  &
        \mathrm{for}~~~|\tau_1-\tau_2| \geq f_{i,j} h \\
        \displaystyle \frac{\tau_1 + \tau_2 + \sqrt{2f_{i,j}^2 h^2-(\tau_1-\tau_2)^2}}{2}, &
        \mathrm{for}~~~|\tau_1-\tau_2| < f_{i,j} h 
    \end{cases},
\end{align}
until convergent, with $\tau_1:= \min(\tau_{i\pm 1,j})$ and $\tau_2:= \min(\tau_{i,j\pm 1}))$ for simplicity. 
\begin{figure}
    \centering
    \includegraphics[width=\columnwidth]{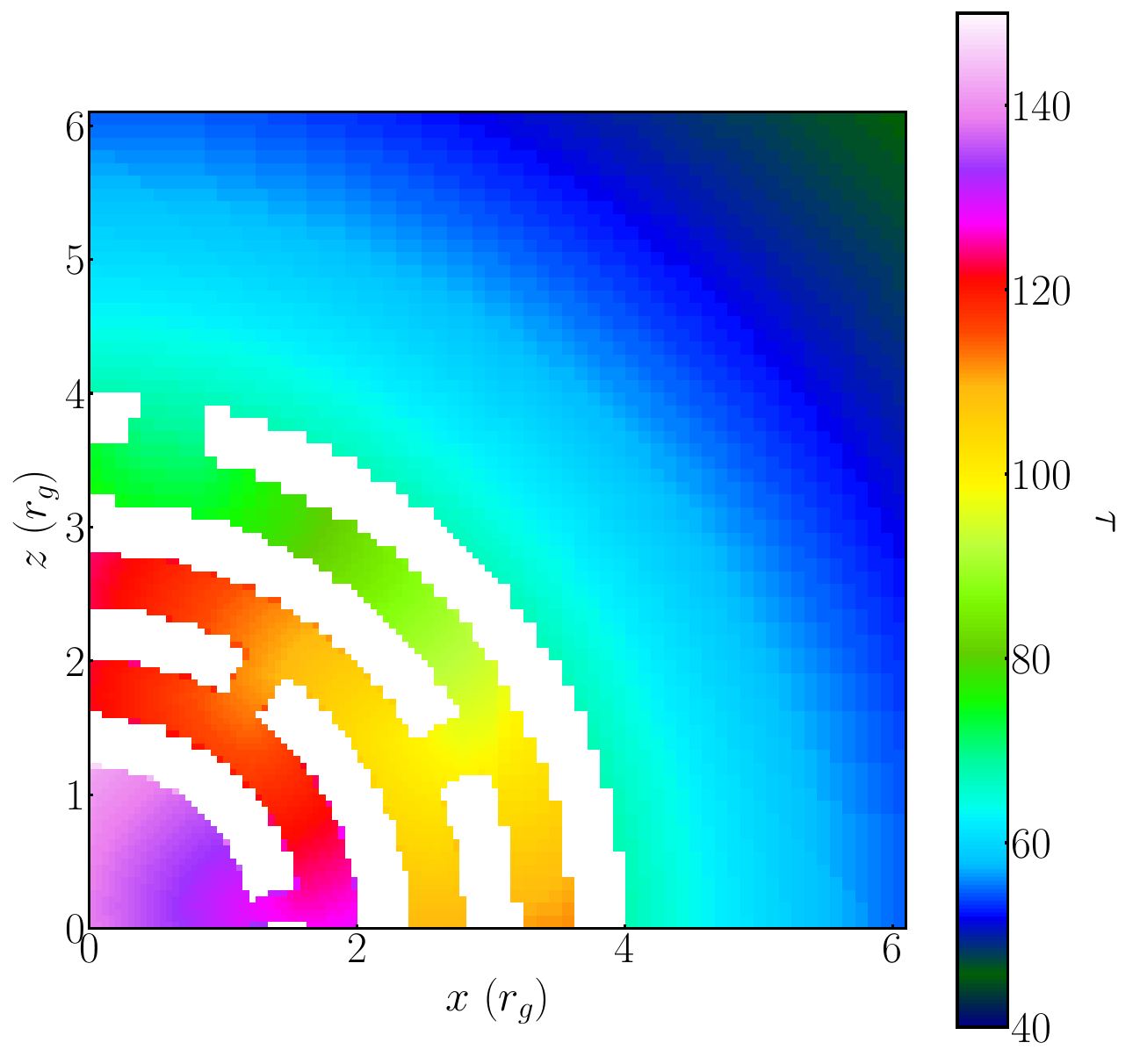}
    \caption{The optical depth obtained by the fast sweeping method in the test problem. The white patches in the snapshot represent the high-density ring regions.
    }
    \label{fig_eikonal}
\end{figure}

To verify our fast sweeping solver, we designed a test problem using a computational domain identical to that in Sec.~\ref{sec3.2}. The density $\rho$ follows Eq.\eqref{eq:profile} for $r \geq 4 r_g$, remaining constant at $\rho=\rho(r=4r_g)$ for $r < 4 r_g$, with parameters $M=10^6 M_\odot$ and $\kappa = 0.35\,{\rm cm}^2/{\rm g}$.
Additionally, we assigned high densities to four rings at $r/r_g$ in the intervals [1.2, 1.6], [2, 2.4], [2.8, 3], and [3.6, 4], each with small random gaps in different directions, creating a maze-like structure.
According to equations~\eqref{eq:tau_est} and \eqref{eq:r_ps}, the optical depth should be roughly $\tau \approx 133\times (r/r_g)^{-1/2}$ for $r>4 r_g$, increasing toward the center along the maze.
Figure \ref{fig_eikonal} displays the results from our fast sweeping solver, which accurately reproduces the analytical solution for $r > 4 r_g$.
For $r<4 r_g$, the white patches with very high optical depth correspond to the high-density ring zones, and the optical depth increases toward the center along the maze, confirming the effectiveness of our solver.

\section{Formulation and Numerical Method} \label{A2}
We implemented general-relativistic viscous hydrodynamics using a simplified version of the Israel-Stewart formulation~\citep{Israel1979a}.
Following previous works~\citep{shibata2017b,fujibayashi2018a}, the stress-energy tensor of the viscous fluid $T_{ab}$ is written as
$T_{ab} = \rho h u_a u_b + P g_{ab} - \rho h \nu_\mathrm{vis} \tau^0_{ab}$,
where $h:= 1 + \varepsilon + P/\rho$ is the specific enthalpy, $\varepsilon$ is the specific internal energy, $u^a$ is the four-velocity, $g_{ab}$ is the spacetime metric, and $\tau^0_{ab}$ is the viscous tensor.
In the Israel-Stewart prescription, the evolution equation for the viscous tensor can be reformulated as
$\mathcal{L}_u \tau_{ab} = - \zeta \tau^0_{ab}$,
where $\mathcal{L}_u$ is the Lie derivative with respect to the four-velocity $u^a$, $\zeta$ is a non-constant coefficient that is the inverse of the relaxation time for $\tau^0_{ab}$ to approach the shear tensor $\sigma_{ab}:= \mathcal{L}_u h_{ab}$,
$h_{ab}:= g_{ab} + u_a u_b$, and $\tau_{ab}:=\tau^0_{ab}+\zeta h_{ab}$.
We refer readers to \cite{shibata2017b} for a more detailed derivation.

%Based on the Valencia formulation, 
The viscous hydrodynamics equations can be written in the following conservative form
\begin{align}
    \del_t \mathbf{q} 
    + \frac{1}{\sqrt{\hat\gamma}} \del_i \left( \sqrt{\hat\gamma}\mathbf{f^i}_{\rm ideal} \right)
    + \frac{1}{\sqrt{\hat\gamma}} \del_i \left( \sqrt{\hat\gamma}\mathbf{f^i}_{\rm vis} \right)
    = \mathbf{s},
\end{align}
with the conservative variables $\mathbf{q}:=(q_D, q_{S_i}, q_E, q_{\tau_{ij}})$ defined as
\begin{align}
    \mathbf{q} := \psi^6
    \begin{pmatrix}
        D \\
        S_i \\
        E \\
        D \tau_{ij}
    \end{pmatrix} = \psi^6
    \begin{pmatrix}
        \rho w \\
        \rho h w u_i - \rho h \nu_\mathrm{vis} \tau^0_{ij} \frac{\bar{u}^j}{w} \\
        \rho h w^2 - P - \rho h \nu_\mathrm{vis}\tau^0_{ij} \frac{\bar{u}^i \bar{u}^j}{w^2} \\
        \rho w \tau_{ij}
    \end{pmatrix},
\end{align}
where $w:= \alpha u^t$ is the Lorentz factor measured by an Eulerian observer, $\alpha$ is the lapse function, and $\bar u^i := \gamma^{ij} u_j$ with $\gamma_{ij}$ being the three-dimensional metric. $\hat \gamma$ is the determinant of the flat metric in the chosen coordinates and $\psi^6=\sqrt{\mathrm{det}(\gamma_{ij})/\hat \gamma}$.
We follow~\cite{Takamoto:2011wi} to split the flux term $\mathbf{f^i} = \left( (f_D)^i, (f_{S_j})^i, (f_E)^i , (f_{\tau_{jk}})^i\right)$ into the ideal fluid part $\mathbf{f^i}_{\rm ideal}$ and the remaining viscous part $\mathbf{f^i}_{\rm vis}$ containing viscosity $\nu_\mathrm{vis}$, which are written as
\begin{align}
    \mathbf{f^i}_{\rm ideal} &= \psi^6
    \begin{pmatrix}
        D \bar v^i \\
        (\rho h w u_j) \bar v^i + \alpha P \delta_j{}^i \\
        (\rho h w^2 - P) \bar v^i + \alpha P \left( \bar v^i + \beta^i \right) \\
        D \tau_{jk} \bar v^i
    \end{pmatrix}, \\
    ~~%\mathrm{and}~~
    \mathbf{f^i}_{\rm vis} &= \alpha \psi^6
    \begin{pmatrix}
        0 \\
        - \rho h \nu_\mathrm{vis} {\tau^0}{}^i{}_j \\
        - \rho h \nu_\mathrm{vis} {\tau^0}{}^i{}_j \frac{\bar u^j}{w}\\
        0, 
    \end{pmatrix}
\end{align}
where $\bar v^i := - \beta^i + \bar u^i / u^t$ denotes the three velocity in the laboratory frame\footnote{There is a typo in Eq.~(20) of \cite{Lam:2025pmz} where $\bar v^i$ should be replaced be $\bar v^i/\alpha$ in the definition of flux term.} and $\beta^i$ is the shift vector.
While we do not strictly separate the inviscid and dissipative steps as in \cite{Takamoto:2011wi}, this flux splitting allows us to use the Riemann solver for the ideal fluid separated from the viscous contribution.
We adopted the Total Variation Diminishing Lax-Friedrichs (TVDLF) method for $(D, S_j, E)$ and the upwind method for $\tau_{jk}$ in $\mathbf{f^i_{\rm ideal}}$.
For $\mathbf{f}^i_{\rm vis}$, we simply take the average of the left and right states at the cell boundary.

The source term $\mathbf{s}:=(s_D, s_{S_i}, s_E, s_{\tau_{ij}})$ in axial symmetry is given by
\begin{align}
    s_D &= 0,~~s_\varphi=0, \\
    s_x &= - S_0 \del_x \alpha + S_j \del_x \beta^j
    - \frac{1}{2}\alpha \psi^6 S_{jk} \del_x \gamma^{jk} \nonumber\\
    &+ \frac{(f_{S_\varphi, {\rm ideal}})^\varphi+(f_{S_\varphi, {\rm vis}})^\varphi}{x},\\
    %s_\varphi &= 0, \\
    s_z &= - S_0 \del_z \alpha + S_j \del_z \beta^j
    - \frac{1}{2}\alpha \psi^6 S_{jk} \del_z \gamma^{jk}, \\
    s_E &= \alpha \psi^6 S_{ij} K^{ij} - \gamma^{ij} S_i \del_j \alpha, \\
    s_{\tau_{ij}} &= -\psi^6 D \left[  \frac{\alpha}{w} \zeta \tau_{0ij} + \tau_{ik} \del_j \bar v^k + \tau_{jk} \del_i \bar v^k  \right], 
\end{align}
where we assumed to use cylindrical coordinates $(x, z, \varphi)$.

In \texttt{SACRA-2D} with adaptive time-subcycling, the time step is given by $\Delta t^{(l)} = c_{\rm CFL} \Delta x^{(l)} / c$ for each refinement level $l$, where $c_{\rm CFL}$ is the Courant–Friedrichs–Lewy factor. However, in the Israel-Stewart formulation, the evolution of viscous hydrodynamics involves stiff source terms that limit the numerical time step $\Delta t$ by the relaxation time $\tau_{\rm relax} = \zeta^{-1}$ with $\zeta = 2/t_\mathrm{g}$ chosen in this work. This causes issues on the coarse grid when $\Delta t > \tau_{\rm relax}$. Although we can avoid this problem by setting $\Delta t < \tau_{\rm relax}$ at all levels and explicitly evolving the equations, this greatly increases computational costs. To circumvent the time-step restriction, an implicit scheme is necessary for stability. We implemented an implicit-explicit (IMEX) Runge-Kutta (RK) scheme, specifically IMEX42L(4,4,2) scheme (\citealt{Izquierdo:2022eaz}), which treats flux terms $\mathbf{f^i}$ and source terms $(s_D, s_{S_i}, s_E)$ explicitly, and the source term $s_{\tau_{ij}}$ implicitly. Since $s_{\tau_{ij}}$ depends linearly on $q_{\tau_{ij}}$, updating $q_{\tau_{ij}}$ implicitly reduces to inverting a $6 \times 6$ matrix inversion.
To simplify further, only $q_{\tau_{ij}} := \psi^6 D \tau_{ij}$ and $q_D :=\psi^6 D$ within $s_{\tau_{ij}}$ are handled implicitly, while the velocity $\bar v^i$ and Lorentz factor $w$ from the prior RK stage are used in the implicit step. This approximation is justified because the dissipative correction to velocity $\bar v^i$ by the dissipative term is expected to be small within the fluid’s characteristic timescale, which is much shorter than $\Delta t = c_{\rm CFL} x / c$. We adopted a constant $c_{\rm CFL} = 0.5$ throughout this study. For boundary interpolation and prolongation at refinement boundary, we used primitive variables $(\rho, u_i, \varepsilon, P, \tau, \tau^0_{ij})$, with $\tau$ denoting the optical depth.

\section{Reproducing Bondi solution}\label{A3}

\begin{figure*}
    \centering
    \includegraphics[width=\textwidth]{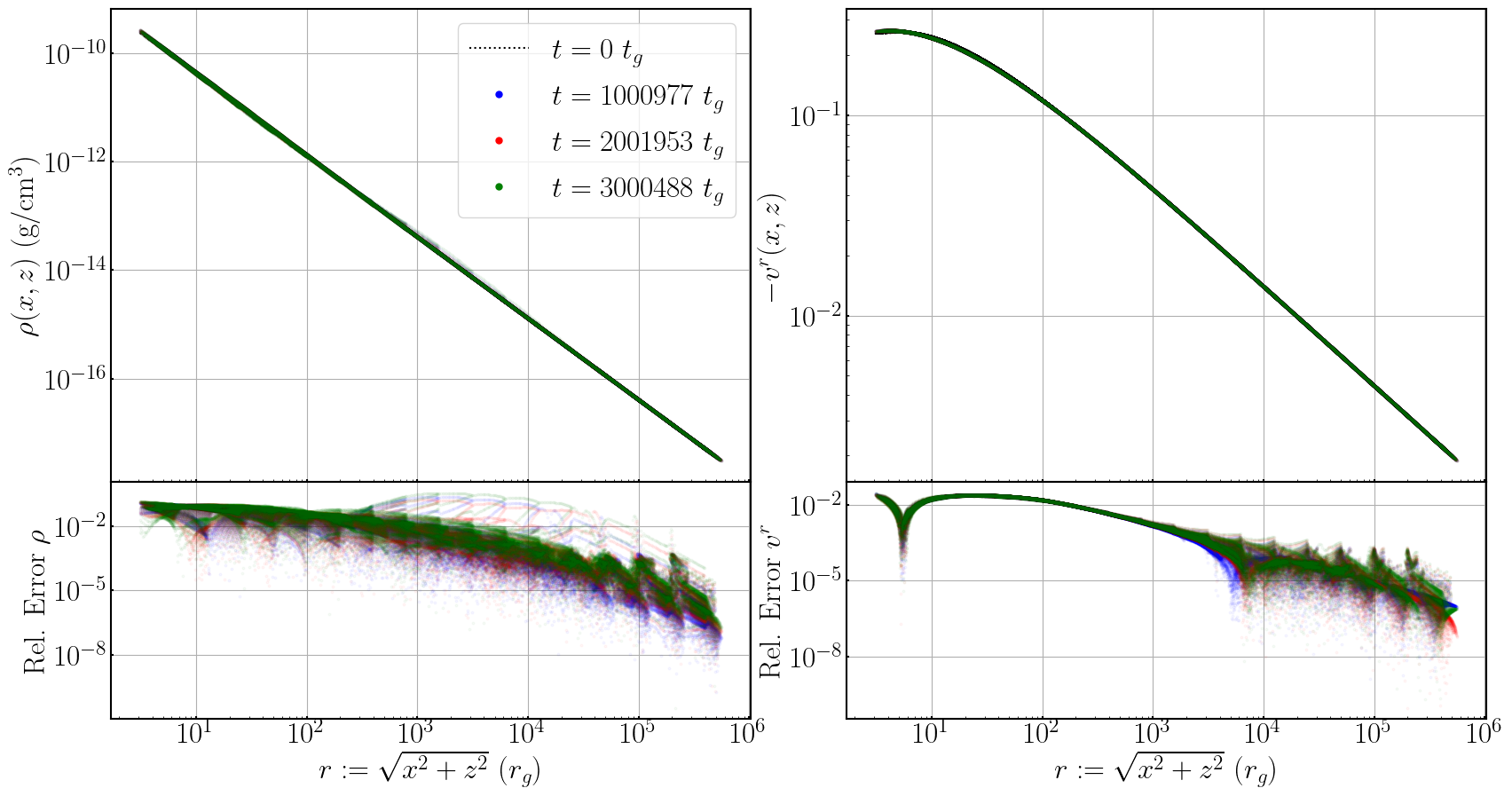}
    \caption{Results of a Bondi flow test. The profile of rest-mass density $\rho$ (left) and radial velocity $v^r$ (right) of Bondi flow at $t=10^6~t_\mathrm{g}$ (blue),
    $2\times 10^6~t_\mathrm{g}$ (red), and $3\times 10^6~t_\mathrm{g}$ (green) as functions of radial coordinate $r$. 
    The dotted line shows the initial profile at $t=0~t_\mathrm{g}$.
    The bottom panels show the relative error of $\rho(t)$ and $v^r(t)$.
    }
    \label{fig_app}
\end{figure*}
\begin{figure}
    \centering
    \includegraphics[width=\columnwidth]{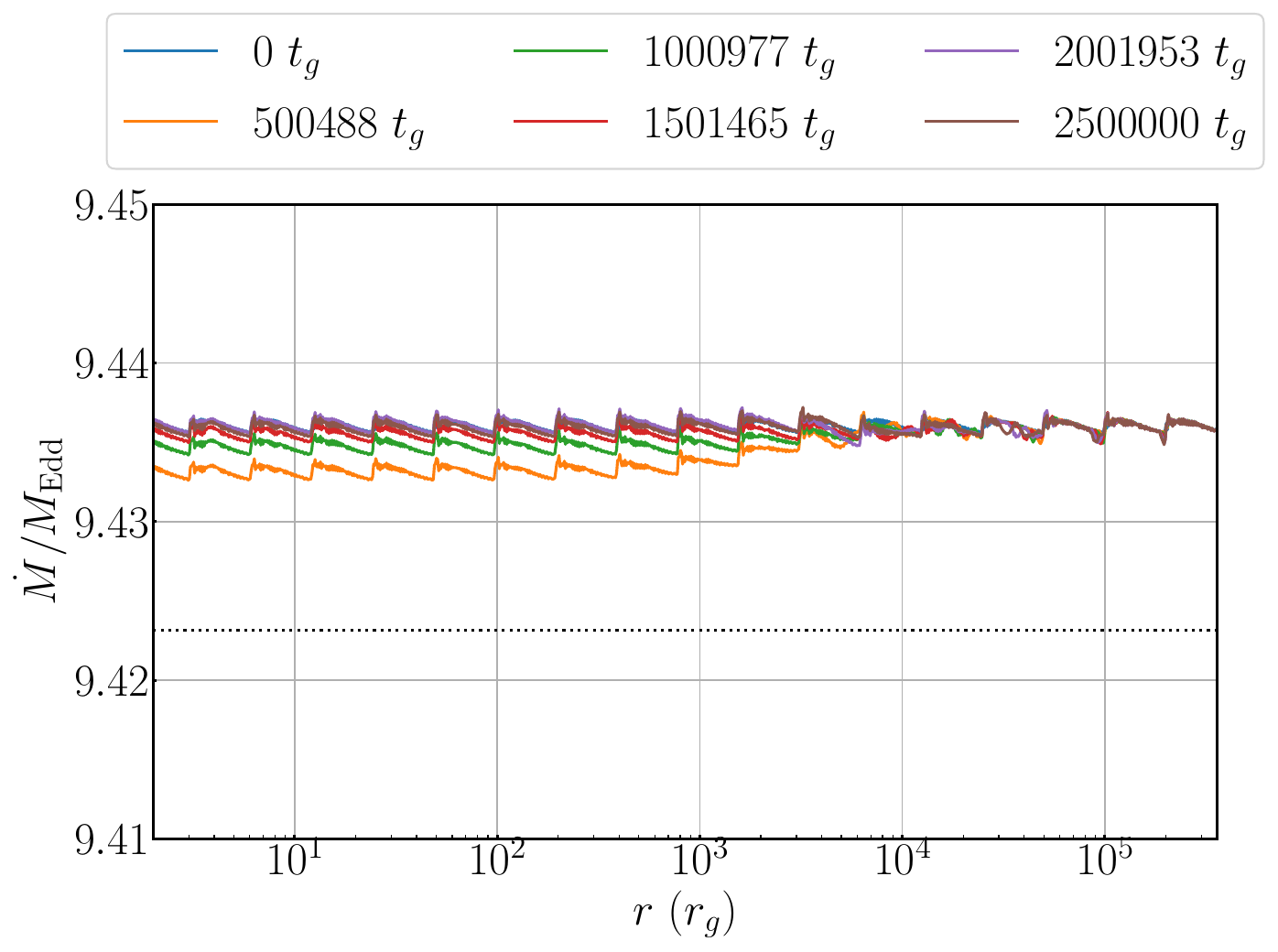}
    \caption{The solid lines show the mass accretion rates in units of $\dot M_\mathrm{Edd}$ as functions of the radius at selected time slices for the Bondi flow test.
    The black dotted line indicates $\dot M = a^3 /G$.
    }
    \label{fig_app2}
\end{figure}

In this work, a key requirement for a numerical hydrodynamics code is that it can reproduce and preserve Bondi solutions for sufficiently long time $\gtrsim 10^6t_\mathrm{g}$ stably and accurately. Here, we show that it is feasible by our code~\citep{Lam:2025pmz}. The grid structure and the sonic point are the same as those employed in the science run for viscous hydrodynamics simulations. 

The top two panels of Figure~\ref{fig_app} show the rest-mass density and radial velocity as functions of the radius in the equatorial plane. The points and solid curves are the numerical results at $t\approx 1$, 2, and $3 \times 10^6 t_\mathrm{g}$ and exact solutions, respectively. It is found that the code can preserve the stationary solution for such a long timescale. 

The bottom two panels of Figure~\ref{fig_app} display the relative error in the rest-mass density and radial velocity as functions of radius. This illustrates that the Bondi solution is preserved even at a late time of $t \approx 3\times 10^6\,t_\mathrm{g}$, with the relative errors within $\sim 10\%$.

Figure~\ref{fig_app2} shows the mass accretion rates as a function of radius at different selected time slices. This illustrates that the mass accretion rates remain approximately constant at late times of $t =2.5\times 10^6\,t_\mathrm{g}$, regardless of the radius, with $\lesssim 0.2\%$ small numerical deviation from the approximated injection rate $\dot M = a^3/G$.

\section{Dependence of the viscous heating rate on the matter profile}\label{A4}

To understand the dependence of the viscous heating rate on the mass and profile of the envelope, we consider equilibrium states of tori around Schwarzschild black holes, assuming that the specific angular momentum is written as $j=A\Omega^{-p}$ where $A$ and $p$ are constants~\citep{Shibata2007sep}. With this choice, $\Omega$ approximately behaves as $\propto R^{-2/(1+p)}$ and for $p\rightarrow 1/3$ and $0$, $\Omega$ approaches the Kepler profile and $j=$constant, respectively. For smaller values of $p$, the torus becomes more geometrically thick. With this profile, the first integral of the Euler equation is written as~\citep{Shibata2007sep}
\begin{eqnarray}
{h \over u^t} + {A \over 1-p}\Omega^{1-p} = C,
\end{eqnarray}
where $C$ is a constant. 

The viscous heating rate is calculated by
\begin{eqnarray}
\dot E_\mathrm{vis}={1 \over 2}\int \nu_\mathrm{vis} \rho \sigma_{\alpha\beta}\sigma^{\alpha\beta} dV,\label{eqD}
\end{eqnarray}
where $\sigma_{\alpha\beta}=\nabla_\alpha u_\beta + \nabla_\beta u_\alpha$ with $\nabla_\alpha$ being the covariant derivative with respect to the spacetime metric. For $\nu_\mathrm{vis}$, we use equation~\eqref{eq27}, by which the viscous heating rate is approximately rewritten to
\begin{eqnarray}
\dot E_\mathrm{vis}\approx \alpha_\mathrm{vis}\int \rho c_\mathrm{s}^2 \Omega \left({\partial \ln \Omega \over \partial \ln R} \right)^2 dV.\label{eq28}
\end{eqnarray}
Hence, the order of the magnitude of $\dot E_\mathrm{vis}$ is written as $\dot E_\mathrm{vis} \sim \alpha_\mathrm{vis} \beta_\Omega^2M_\mathrm{b} c_\mathrm{s}^2 \Omega$ where $M_\mathrm{b}$ is the rest mass of the torus. Note that the results of equations~\eqref{eq28} agree with those of equation~\eqref{eqD} within $\sim 10\%$ error for the tori considered in this Appendix. 

The torus profiles are determined by giving the inner and outer edges of the torus on the equatorial plane, for which the radii are referred to as $r_1$ (inner) and $r_2$ (outer). By giving $r_1$ and $r_2$, $A$ and $C$ are determined. A polytropic equation of state, $P=K\rho^\Gamma$, is employed with $\Gamma=4/3$, assuming that the torus is dense and radiation-pressure dominant. By varying the polytropic constant $K$, $M_\mathrm{b}$ is varied. Assuming that the viscous heating rate is equal to the Eddington luminosity with the surface temperature of $T \sim 5000$\,K, $r_2$ is set to be $4 \times 10^5$, $1.2\times 10^5$ and $4 \times 10^4 r_\mathrm{g}$ for $M=10^5$, $10^6$ and $10^7M_\odot$, respectively. We determine $M_\mathrm{b}$ by the condition of $\dot E_\mathrm{vis}=L_\mathrm{Edd}$ for given values of $p$ and $r_1$. We note that the torus is non-spherical and the Eddington luminosity should be determined taking into account the non-spherical morphology, but in this work, we approximately employ the expression for the Eddington luminosity in spherical symmetry. 

\begin{figure}
    \centering
    \includegraphics[width=\columnwidth]{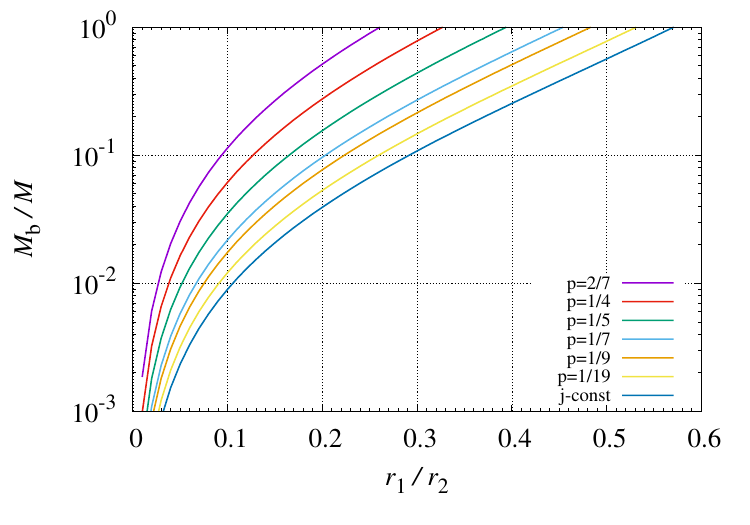}
    \includegraphics[width=\columnwidth]{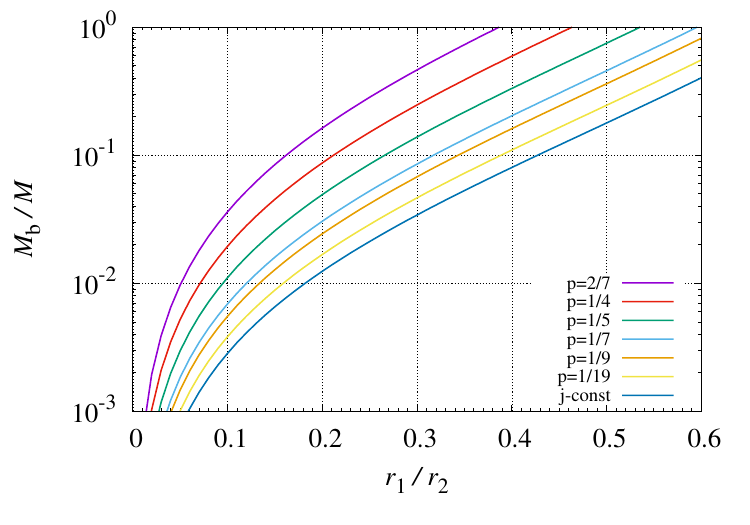}    \caption{$M_\mathrm{b}/M$ as a function of $r_1/r_2$ for several values of $p$ with $M=10^5M_\odot$ (upper panel) and $10^7M_\odot$ (lower panel). We note for $M_\mathrm{b}/M$ close to unity, the self-gravity of the torus should be important and the curves plotted here are not quantitatively accurate. 
    }
    \label{fig_app4}
\end{figure}

Figure~\ref{fig_app4} shows $M_\mathrm{b}/M$ as a function of $r_1/r_2$ for several values of $p$ with $M=10^5M_\odot$ (upper panel) and $10^7M_\odot$ (lower panel). $\alpha_\mathrm{vis}=0.05$ is chosen in computing the viscous heating rate. For each curve, $\dot E_\mathrm{vis}=L_\mathrm{Edd}$ is satisfied, and for the larger $M_\mathrm{b}$ side above each curve, the torus should not exist because of $\dot E_\mathrm{vis} > L_\mathrm{Edd}$. For larger black hole masses, the curves are systematically shifted to the lower $M_\mathrm{b}/M$ side (compare the upper and lower panels of Fig.~\ref{fig_app4}). 

It is found that the larger torus mass is possible for smaller values of $p$ (i.e., for angular velocity profiles close to the Kepler one) and for the tori located further away from the black hole. This suggests that when the envelope matter is predominantly located near the photosphere, a high mass envelope with $M_\mathrm{b} > M$ may be possible. On the other hand, if the torus is compact, high masses would be prohibited due to high viscous heating. Thus, it is expected that with the increase of the envelope mass, the matter near the black hole is viscously heated, goes away from the inner region through the convective motion, and may be distributed to the region near the photosphere. As a result, the mass infall rate near the black hole would be suppressed even for high values of $M_\mathrm{b}/M$. For further increases in the envelope mass, the viscous heating rate will eventually exceed the Eddington luminosity, and the mass increase will then stop. In such a situation, an inflow-outflow structure may be an outcome. This point will be studied in our future work. 

%\newpage
\label{lastpage}
\end{document}